\documentstyle[preprint,eqsecnum,aps]{revtex}
\newcommand{\QO}{({\bf q},\omega )}

\newcommand{\eq}{\begin{equation}}
\newcommand{\en}{\end{equation}}
\draft
\input epsf
\begin{document}
%%%%%%%%%%%%%%%%%% TITLE %%%%%%%%%%%%%%%%%%%%%%%%%%%%%%%%%%
\title{Metastable Dynamics above the Glass Transition}
\author{Joonhyun Yeo and Gene F. Mazenko}
\address{
The James Franck Institute 
and Department of Physics \\ 
The University of Chicago,
Chicago, Illinois  60637}
\date{\today}
\maketitle
\begin{abstract}
The element of metastability is incorporated in the fluctuating
nonlinear hydrodynamic description of the mode coupling theory (MCT)
of the liquid-glass transition. This is achieved through the
introduction of the defect density variable $n$ into the set of 
slow variables with the mass density $\rho$ and the momentum 
density ${\bf g}$. As a first approximation, we consider 
the case where motions associated with $n$ are much slower than
those associated with $\rho$. Self-consistently, assuming
one is near a critical surface in the MCT sense,
we find that the observed slowing down of the dynamics corresponds 
to a certain limit of a very shallow metastable well
and a weak coupling between $\rho$ and $n$. The metastability parameters
as well as the exponents describing the observed sequence of time 
relaxations are given as smooth functions of the temperature 
without any evidence for a special temperature. 
We then investigate the case where the defect dynamics
is included. We find that the slowing down of the 
dynamics corresponds to the system arranging itself such that 
the kinetic coefficient $\gamma_v$ governing the diffusion of 
the defects approaches from above a small 
temperature-dependent value $\gamma^c_v$. 
\end{abstract}
%%%%%%%%%%%%%%%%%%%%%%%%%%%%%%%%%%%%%%%%%%%%%%%%%%%%%%%%%%%%%%%%%%
\narrowtext
\section{introduction}
\label{sec:intro}
While there has been considerable recent progress in the development
of the theory of relaxation near the liquid-glass transition, these 
theories have ignored one of the fundamental defining qualities
of the problem: These systems are metastable. In this paper,
we introduce the element of metastability into a framework 
that connects with mode coupling theory (MCT) \cite{mct:gen}.
We find a theoretical picture which is closer to
the observed experimental picture than conventional MCT, since
it is freed from the idea of a sharp transition temperature $T_0$.

The dynamics of supercooled liquids is characterized by very long 
relaxation times and a rapid increase of the viscosity
with decreasing temperature. Thermodynamic quantities such as
specific heat,
compressibility, etc. show a change when the temperature reaches 
the so-called calorimetric glass transition 
temperature $T_g$ which is defined
somewhat arbitrarilty by the temperature where the viscosity
reaches $\sim 10^{15}$ poise. Recent interests, however,
have focused on the study of the 
relaxation of supercooled liquids at temperatures well
above $T_g$. The reason for this focus has been mainly due to
the discovery of the
existence of a very elaborate sequence of time
relaxations spread over many decades of time. 
This sequence has been observed in a series of experiments including 
light scattering \cite{light,light1}, neutron scattering \cite{neutron}
and dielectric measurements \cite{dielec,nagel}.
As depicted for the 
density auto-correlation function $\phi (t)$ in Fig.~1, 
this sequence 
of time relaxations is a
three-step process with characterizing exponents, $a, b$ and 
$\beta$. After a fast microscopic process, $\phi (t)$
decays as a power law, $f+A_1 t^{-a}$ followed by the
so-called von-Schweidler relaxation, $f-A_2 t^b$. For longer times
$t>\tau_{\alpha}$, $\phi (t)$ decays with a stretched exponential,
$\exp \{ -(t/\tau_{\alpha})^\beta\}$, $\beta <1$. 

The mode coupling theory has been very successful
in explaining this sequence of time relaxations.
As originally formulated by
Leutheusser \cite{leut}, MCT is based on
the nonlinear feedback mechanism where the renormalized 
viscosity is expressed in terms of the integral of
the quadratic term in
density auto-correlation functions. 
If we ignore the coupling of the density to the energy density,
the Fourier-Laplace transform
of $\phi (t)$ can be written in the generalized-hydrodynamics form
\cite{dmrt}
\eq
\phi (z)=\frac{z+i\Gamma (z)}{z^2-\Omega^2_0
+iz\Gamma (z)}, \label{1}
\en
where $\Omega_0$ is a microscopic `phonon' frequency and
$\Gamma (z)$ is the renormalized viscosity and the wavenumber
dependence has been suppressed. In MCT \cite{fn:1} it is assumed
that $\Gamma (z)$ can be expressed in the form
\eq
\Gamma (z)=d_0 + \Omega^2_0\int^\infty_0 \; dt \:
e^{izt}\:\sum^{N}_{n=1} c_n \phi^n (t) ,\label{Gamma}
\en
where $d_0$ is the bare viscosity. 
The original theory due to Leutheusser \cite{leut} corresponds to
all $c_n =0$ except $c_2$. This theory does not lead to the
von-Schweidler regime and does not give stretched dynamics. Instead,
one finds the exponential decay ($\beta =1$) following the power 
law decay regime. 
The importance of the term linear in $\phi (t)$ ($c_1$) in 
$\Gamma (z)$ was emphasized by G\"otze \cite{Gotze:lin}
as a sufficient condition for obtaining the stretched exponential
relaxation and von-Schweidler regime. 

It was first noticed in Ref.~\onlinecite{dmrt} that MCT 
can be understood in terms of the fluctuating nonlinear
hydrodynamics (FNH). This analysis was then formulated in detail by Das
and Mazenko \cite{dm}. One can then
use a well-developed field-theoretic methods \cite{MSR,MSR1}
to determine $\Gamma (z)$ perturbatively
for almost any choice of the free energy.
In this formalism, it can be understood \cite{kim}, for example,
how the linear term in $\Gamma (z)$ is generated. G\"otze had 
originally introduced the $c_1$ term on phenomenological
grounds \cite{Gotze:lin}. 

Despite its success, the current status of MCT is not without
questions and controversies. First of all, the scaling picture 
arising from the experiments of Dixon et al.~\cite{nagel} shows 
universal features beyond those suggested by MCT. In their dielectric 
measurements, they were able to find a scaling curve that fits all
the data obtained over 13 decades of frequency range for various 
glass forming liquids and for different temperatures. This
beautiful result has not yet found an explanation within MCT.
As indicated by Kim and Mazenko \cite{km}, the data of 
Ref.~\onlinecite{nagel}
also implies a universal relationship between the von-Schweidler
exponent $b$ and the stretching exponent $\beta$:
$(1+b)/(1+\beta )=$universal constant. Although this result can be
made compatible with MCT \cite{km}, it requires the introduction of 
additional arbitrary constraints.

Another important question associated with MCT concerns
the temperature dependence of the theory. 
Conventional MCT \cite{mct:gen,leut,Gotze:lin,Gotze:1} 
views the glass transition as a sharp 
ergodic-nonergodic transition as the temperature $T$
approaches the ideal glass transition temperature $T_0$
well above $T_g$. According to conventional MCT, the relaxation
sequence occurs only near $T\sim T_0$ and 
in particular the von-Schweidler relaxation and the 
stretching are confined to the region $T>T_0$. This 
sharp temperature dependence is clearly in
contrast with the universal behavior of Dixon et al.~whose data cover 
any reasonable choice of $T_0$. Also, as shown in other
recent experiments \cite{no:t0}, the exisistence of 
a well-defined
temperature $T_0$ well above $T_g$ is very doubtful.
In this formulation of MCT, the exponents governing the time 
relaxations are temperature independent just as those involved in
a second order phase transition. This is again in disagreement with
recent experiments. In particular, Dixon et al.~finds $\beta$ weakly
temperature dependent, which means, by virtue of the universal
relationship mentioned above, that $b$ also depends
on temperature. Since MCT predicts \cite{Gotze:1}
that the exponents $a$ and $b$
are related through a temperature-independent relation, this
also implies that $a$ is temperature dependent.

Das and Mazenko \cite{dm}, in their study
of FNH of dense fluids, discovered that there exists an important
nonhydrodynamic correction that cuts off the sharp nature of the 
ideal glass transition. This cutoff corresponds to
an extra contribution
$\gamma (z)$ in Eq.~(\ref{1}):
\eq
\phi (z)=\frac{z+i\Gamma (z)}{z^2-\Omega^2_0
+i\Gamma (z)[z+i\gamma (z)]} \label{cutoff}
\en
Although it is now widely believed that
the sharp transition is indeed smeared by this cutoff mechanism,
the question of the exisistence of $T_0$ still remains. 
There exist recent efforts \cite{ext:mct} to reconcile the 
discrepancies between experiments and MCT, which includes
the cutoff effect into
conventional MCT but still assumes a rounded transition 
around $T\sim T_0$. This attempt, however, ends up with adjusting 
as many as seven independent parameters to fit experimental data.
A simpler interpretation of the situation is that 
MCT, despite all the successes, needs to be reformulated at a
more fundamental level such that it is not tied to the notion
of a sharp transition temperature.

In Ref.~\onlinecite{my:2}, we proposed a model where
defects and 
metastabilty play an important role in the glass transition in 
an attempt to reformulate MCT such that it is compatible with
experiments. Although this new theory does not provide any clear
explanation for the scaling result of Ref.~\onlinecite{nagel},
it does result in a smooth temperature variation without any 
indication of a special temperature $T_0$. 
The various exponents, according to the model, are weak functions of
temperature.
In this paper, our main focus is to elaborate on
the detailed construction of the model, which was
omitted in Ref.~\onlinecite{my:2}, and to further 
elaborate on the dynamics of the defects which we treated using
a simplifying assumption in the
previous work. 

The model introduced in 
Ref.~\onlinecite{my:2} is based on the introduction of the 
defect variable $n$ into
the set of slow variables consisting of the mass density
$\rho$, and the momentum density ${\bf g}$,
in the FNH description of MCT \cite{dm}. 
The introduction of the defect variable would be required in 
a rigorous hydrodynamic description of crystalline solids,
along with the Nambu-Goldstone modes associated with
broken translational symmetry \cite{mpp}. Although there is no broken
continuous symmetry involved in the glass transition, 
we consider the situation 
where this variable plays a role in the transition \cite{cohen}.
The best analogy to the present case 
would be the situation where
one considers an order parameter
in the disordered state. 
We do not need a
microscopic definition for the defect variable here.
The only information we need here is that 
the defect density has the usual Poisson bracket relations of a 
scalar variable with the momentum density ${\bf g}$,
that they are metastable and that they interact weakly
with the mass density.

One of the key assumptions in the model
is that motions associated with
the defect density $n$ have a very long time scale
compared to that of density fluctuation.
This is realized in the
model via an explicit double well potential 
$h(n)$ for the defect variable 
with the metastable defect density $\bar{n}$ associated with 
the minimum of the higher well. 
A very small diffusion coefficient $\Gamma_v$ for $n$
results from a rolling around in a shallow metastable well.
We find that
the coupling between the mass density and the defect variables
slows down the defect motion further.
The coupling also enables the slow dynamics of defects 
to influence the density dynamics.
Over the significant time period where the defect is trapped
within the metastable well, the
defect auto-correlation function $\psi(t)$ 
can be regarded as a constant 
while the density auto-correlation function $\phi(t)$ 
displays the relaxation sequence. This is
essentially the basic mechanism discussed in
Ref.~\onlinecite{kim} that generates the linear term 
in the mode coupling integral. Since one variable 
(in this case the defect variable)
is extremely slow compared to the other, we can replace
a term like $\psi(t)\phi(t)$ in $\Gamma (z)$ by a constant times
$\phi(t)$. Under this assumption, we find that
the observed stretched dynamics corresponds to a certain
self-consistent limit of weak coupling and low activation
barrier for the defect. 
As will be discussed below, this limit corresponds to
the situation where the coupling energy is weak enough
not to destroy the metastable defects but still strong 
enough for the slow dynamics of defects
to result in the slowing down of the mass density variable.
In this limit, the parameters describing the 
double-well potential and the coupling, as well as the 
exponents of the relaxation sequence are
self-consistently determined as smooth functions of temperature.  
  
Since the defect variable is diffusive, 
the defect auto correlation
function $\psi(t)$ decays, 
to linear order, as $\sim\exp (-\gamma_v t)$,
where in terms of the bare diffusion coefficient
$\Gamma_v$ for $n$, 
$\gamma_v =h^{\prime\prime}(\bar{n})
\Gamma_v q^2$ at the wavenumber $q$.
Self-consistently we find,
using the previous assumption that $\psi (t)$ is
regarded as a constant, which
corresponds to taking the limit
$\gamma_v\rightarrow 0$,
that $h^{\prime\prime}(\bar{n})$
must be small, {\it i.e.} the metastable wells become very
shallow and broad.
As one moves into the later stage of relaxation, however,
we expect that nonlinear
corrections to $\gamma_v$ become important. Therefore, 
we need to consider the dynamics of the defects and their 
coupling to the density fluctuations by 
including the renormalization of $\gamma_v$ in
the evolution equation for $\psi (t)$. In this case,
$\Gamma (z)$ in Eq.~(\ref{1})
is not represented simply by the equation for
$\phi (t)$ alone as in Eq.~(\ref{Gamma}),
since the $c_1$ term is now written as $\psi (t)
\phi (t)$ with a nonconstant $\psi (t)$. Instead, we obtain
a set of coupled equations for $\phi (t)$ and $\psi (t)$
governing the dynamics of both density and defect 
variables.

We find, through mainly numerical investigations, that
the extended model including the defect dynamics, provides
considerable self-consistent information on the nature of 
the defects near the 
glass transition. In particular, we find that the bare 
value of $\gamma_v$ is significantly restricted
when the observed stretching occurs. In fact the long 
time scale of the defect variable corresponds to the case where
the system arranges itself such that $\gamma_v$ is close to 
some small temperature-dependent value $\gamma^c_v$.
The system seems to pick
out its own defect potential and diffusion coefficient.
We find that, when $\gamma_v\simeq\gamma^c_v$, the time
scale associated with the defects become much longer
than that of the density variable, and thus the basic picture
obtained using the previous simplifying assumption still holds:
The parameters describing the metastable wells and the coupling 
arrange themselves such that the coupling enables the slow
dynamics of defects to effect the density dynamics without
destroying the metastable wells.

We note that the approach we take in this paper is not to
explain how the defects bring the system to the observed slowing
down starting from the microscopic description of the defects.
Instead, we assume that the system arranges itself to be on the 
critical surface associated with MCT. We then investigate 
the conditions that the defect degrees of freedom must satisfy
to be on this critical surface.

The cutoff mechanism of Das and Mazenko \cite{dm}
will eventually influence the very long time dynamics by
generating an exponential decay. 
We note that there exists a mechanism \cite{Sjo,km}
that drives $\gamma (z)$
to a small value so that the observed slowing down is retained.
Although,
in principle we can include this effect into the coupled equations,
we consider here the situation where $\gamma (z)=0$.
Another simplification made in this analysis is
to neglect the wavenumber
dependences of correlation functions. 
In terms of FNH, one can construct a wavenumber dependent model as
shown in Ref.~\onlinecite{Das}.
Since this model is difficult to
analyze numerically, we focus here
on the wavenumber independent model.

In Sec.~\ref{sec:form}, we present a detailed formulation of
FNH of simple fluids including the defect variable. 
We then use the well-known
field-theoretic techniques
to calculate the relevent nonlinear contributions to the 
glass transition.
In Sec.~\ref{sec:cnn1}, the case
where the defect auto-correlation is a constant is considered
in detail
as a first approximation. This will reproduce the results
of Ref.~\onlinecite{my:2}.
The analysis of the full model which
includes the defect dynamics is presented in Sec.~\ref{sec:main}.
In Sec.~\ref{sec:cut}, we discuss the cutoff mechanism and
the temperature dependence of the viscosity in
this formulation. In Appendix, we consider the case of a 
general potential
$h(n)$ without assuming the particular double-well form.

\section{FNH with the Defect Density Variable}
\label{sec:form}

In this section, we formulate the FNH of compressible
fluids in detail including the defect
density $n$ as the additional slow variable.
Using well-developed field theoretic methods,
we calculate nonlinear corrections to the density and defect 
autocorrelation functions systematically.
\subsection{Generalized Langevin Equations}
Our starting point is the generalized
Langevin equation
for the set of slow variables, $\psi_\alpha =\rho ({\bf x}), 
g_i ({\bf x}), n({\bf x})$, where
$\rho$ is the mass density and ${\bf g}$ is the momentum
density. Here $\alpha$ labels the type of the field,
the position
${\bf x}$, and the vector label $i$. 
Following the standard procedure described
by Ma and Mazenko \cite{MM}, we have the equation of
motion given by
\eq
\frac{\partial\psi_\alpha}{\partial t}=\bar{V}_\alpha
[\psi ]-\sum_\beta \Gamma_{\alpha\beta}\frac{\delta F}{\delta
\psi_\beta} +\Theta_\alpha,
\en
where  $F$ is the effective Hamiltonian and
$\bar{V}_\alpha$ is the streaming velocity governing
the reversible dynamics and is given by
\eq
\bar{V}_\alpha [\psi ]=\sum_\beta \{\psi_\alpha ,
\psi_\beta \}\frac{\delta F}{\delta\psi_\beta },
\label{stream}
\en
and $\{\psi_\alpha ,\psi_\beta \}$ is the Poisson bracket 
among the slow variables. The dissipative matrix
$\Gamma_{\alpha\beta}$ and the Gaussian noise $\Theta_\alpha$
satisfy
\eq
\langle\Theta_\alpha (t)\Theta_\beta (t^\prime )\rangle
=2k_B T\Gamma_{\alpha\beta}\delta (t-t^\prime ).
\en

The effective Hamiltonian for
$\psi_\alpha$
is given by
\eq
F=F_K +F_u [\delta\rho ]+F_v [\delta\rho, n],
\label{F}
\en
where $F_K$ is the kinetic energy:
\eq
F_K =\int\, d^3 {\bf x}\:\frac{{\bf g}^2 ({\bf x})}{2\rho
({\bf x})},
\label{FK}
\en
$F_u [\delta\rho ]$ is the potential energy for the
density fluctuation, $\delta\rho =\rho -\rho_0$, where 
$\rho_0$ is the average density, and
$F_v [\delta\rho ,n]$ governs the defect density and
its coupling to $\rho$. In general,
$F_u [\delta\rho ]$ can be any local functional of 
$\delta\rho$ and the spatial derivatives of $\delta\rho$.
In particular, one can study the wave-number dependence of the
structure factor by including the spatial derivatives \cite{Das}.
Recently it has been claimed \cite{wavenumber}
that by including the wave number 
dependences one can effectively generate the von-Schweidler and the 
stretching. Analytical treatment in this case, however, is very 
difficult.
In the present case, we consider the simple quadratic
form which corresponds to a wave-number independent structure factor,
\eq
F_u [\delta\rho]=\int\, d^3 {\bf x}\:\frac{A}{2}\Bigl( \delta\rho 
({\bf x}) \Bigr) ^2,
\label{Fu}
\en
where $A$ is the flat inverse susceptibility
and wavenumbers are restricted to values less than a cutoff $\Lambda$.
$F_v [\delta\rho ,n]$ in Eq.~(\ref{F}) is assumed to be of the form:
\eq
F_v [\delta\rho, n]=\int\, d^3 {\bf x}\: \Bigl[ B\delta\rho(
{\bf x})n({\bf x})+h(n({\bf x}))\Bigr] ,
\label{Fv}
\en
where we introduced the simple coupling term through the
coupling constant $B$. 
As discussed in Sec.~\ref{sec:intro}, we construct 
the potential energy $h(n)$
to be a double well potential \cite{fn:2} with
the metastable defect density $\bar{n}$
associated with the higher well.
We parametrize $h(n)$ such that $h(n)$ has three extrema at $
n=0$, $(1-\sigma )\bar{n}$ and $\bar{n}$:
\eq
h^\prime (n)=\epsilon n \bigl( n-(1-\sigma )\bar{n}\bigr)
(n-\bar{n}),
\en
or upon integrating,
\eq
h(n)=\epsilon\bar{n}^4 \Bigl[ \frac{1}{4}(\frac{n}{\bar{n}})^4
-\frac{1}{3}(2-\sigma )(\frac{n}{\bar{n}})^3
+\frac{1}{2}(1-\sigma )(\frac{n}{\bar{n}})^2 \Bigr] .
\label{hhnn}
\en
The parameter $\epsilon$ gives the correct 
dimension for $h(n)$ and describes the overall scale of the
potential energy. 
For small positive $\sigma$, Eq.~(\ref{hhnn}) represents a 
double well potential with the global minimum at $n=0$ and
the metastable minimum at $\bar{n}$.
We note that for small negative $\sigma$, $h(n)$ in the
above representation is also a double well potential.
In fact, $h(n)$ is invariant under $\sigma\rightarrow
\sigma^\prime \equiv -\sigma /(1-\sigma )$ and
$\bar{n}\rightarrow\bar{n}^\prime \equiv (1-\sigma )\bar{n}$.
Thus, in the following analysis, only the absolute value of
$\sigma$ will play a role. We also note that
if $\sigma =0$, then $h(n)$ develops an inflexion point at
$n=\bar{n}$.
It is useful to
define two dimensionless parameters $x$ and $y$
which characterize the scale of the coupling energy
and the potential energy respectively:
\eq
x\equiv \frac{B\rho_0 \bar{n}}{A\rho^2_0 },~~~~~
y\equiv \frac{\epsilon\bar{n}^4 }{A\rho^2_0 }.
\en
The potential and coupling energy are then completely described
by $x,y,\sigma$ and $\bar{n}$. We consider the coupling energy term 
in Eq.~(\ref{Fv})
as a small distortion to the shape of $h(n)$.
In particular, $\bar{n}$ is shifted to $n^{*}[\delta\rho ]$
determined by
\eq
0=\left.\frac{\delta F_v}{\delta n}\right| _{n=n^{*}
[\delta\rho ]}=B\delta\rho +h^\prime (n^{*}[\delta
\rho ]).
\label{n*}
\en
If we expand in powers of $\delta\rho$,
\eq
n^{*}[\delta\rho ]=\bar{n}\Bigl[ 1+a_1 (\frac{\delta\rho}
{\rho_0})+
a_2 (\frac{\delta\rho}{\rho_0} )^2 +\cdots \Bigr] ,
\label{expansion}
\en
and we can easily calculate the coefficients $a_1 , a_2 , \cdots$ 
using Eq.~(\ref{n*}):
\eq
a_1 =-\frac{x}{\sigma y},~~~~a_2 =-
(\frac{1+\sigma }{\sigma})(\frac{x}{\sigma y})^2,~~~\cdots\, .
\en
We will later consider the fluctuation of $n$
around this shifted $n^{*}[\delta\rho ]$: $\delta n=n-n^{*}
[\delta\rho ]$.
Eq.~(\ref{expansion}) indicates that,
for small $|\sigma |$,
the coupling energy $x$ must be sufficiently
small such that
\eq
|a_1|\ll 1,~~~~~~~|a_2|\ll 1
\label{abll1}
\en
in order
to have a sensible expansion.

Turning to the reversible streaming velocity terms in the
Langevin equations, we assume that
the Poisson brackets involving $\rho$ and ${\bf g}$ are 
evaluated in the usual way as in Ref.~\onlinecite{dm}. The new variable
$n$ is a scalar quantity as $\rho$, thus we assume that the 
Poisson brackets for $n$ have the same structure as those for
$\rho$. The only nonvanishing elements involving $n$ are
\begin{eqnarray}
&&\{ n({\bf x}), g^i ({\bf x^\prime})\}=-\nabla^i_{{\bf x}}
[\,\delta ({\bf x-x^\prime })n({\bf x})\, ], \nonumber \\
&&\{ g^{i}({\bf x}), n({\bf x^\prime })\}=\nabla^{i}_
{{\bf x^\prime}}[\,\delta ({\bf x-x^\prime})n({\bf x})\, ].
\end{eqnarray}

Since there are no Poisson brackets relating $\rho$
to $n$ and
$\Gamma_{\rho\beta}=0$, the Langevin equation for $\rho$
is simply the continuity equation,
\eq
\frac{\partial\rho}{\partial t}=-{\bf \nabla\cdot g}
\label{rho}
\en 
The streaming velocity for $g^i$ then follows from 
Eq.~(\ref{stream}):
\begin{eqnarray}
\bar{V}^i_g &=&\int d^3 {\bf y} \Bigl[ \{ g^i ({\bf x}),
\rho ({\bf y})\} \frac{\delta F}{\delta\rho 
({\bf y} )}+\{ g^{i}({\bf x}),n({\bf y} )\}
\frac{\delta F}{\delta n({\bf y} )} 
+\sum_j  \{ g^i ({\bf x}),
g^j ({\bf y} )\}\frac{\delta F}{\delta g_j
({\bf y} )}\Bigr] \nonumber \\
&=& -\rho ({\bf x})\nabla^i_{\bf x}\frac{\delta F}{\delta
\rho ({\bf x})}- n({\bf x})\nabla^i_{\bf x}\frac{\delta F}
{\delta n({\bf x})}-\sum_j\nabla^j_{\bf x}\Bigl(\frac{g_i g_j}
{\rho}({\bf x})\Bigr) . \label{Vig}
\end{eqnarray}
Using the explicit form of the effective Hamiltonian,
Eqs.~(\ref{FK}-\ref{Fv}), $\bar{V}^i_g$ can be expressed
as a functional of $\delta\rho , g^i$ and $n$. In terms 
of the fluctuation, $\delta n=n-n^{*}[\delta\rho ]$,
we have, using Eq.~(\ref{expansion}),
\eq
\bar{V}^{i}_{g}=-\sum_{j}\nabla_{j}(\sigma^{u}_{ij}
+\sigma^{v}_{ij})
-\sum_{j}\nabla_{j}(\frac{g_i g_j}{\rho}),\label{Vig*}
\en
where 
\eq
\sigma^u_{ij}\equiv\delta_{ij}\{ c^2_0 (\delta\rho )
+\frac{1}{2}\widetilde{A}(\delta\rho )^2 \}
\label{sigmau}
\en
is the usual stress tensor
with the definitions,
\begin{eqnarray}
&&c^{2}_{0}\equiv A\rho_{0}(1-
\frac{x^2}{\sigma y}) \\
&&\widetilde{A}\equiv A\Bigl( 1-\frac{x^2}{\sigma y}-
\frac{2(1+\sigma )x^3 }{\sigma^3 y^2}\Bigr) .
\end{eqnarray}
The bare sound 
speed $c^2_0$ and the inverse susceptibility $\widetilde{A}$
are modified due to the coupling. But the terms depending on the
coupling constant are even smaller than $|a_1|$ and $|a_2|$
by a factor of $|x|$. Thus we may neglect them and regain
the usual relations: 
\eq
c^2_0 =A\rho_0, ~~~~~~\widetilde{A}=A.
\en
The new $n$-dependent
terms in Eq.~(\ref{Vig*}) are given by
\eq
\sigma^v_{ij}\equiv \delta_{ij}\, (A\rho^2_0) \Bigl[
(x+\sigma y)(\frac{\delta n}
{\bar{n}})
-2x(\frac{1+\sigma }{\sigma })
(\frac{\delta\rho}{\rho_0} )(\frac{\delta n}{\bar{n}})
+y(1+\frac{3\sigma}
{2})(\frac{\delta n}{\bar{n}})^2 \Bigr] .
\label{sigmav}
\en
For our purpose, we need only to keep upto quadratic 
terms in the fluctuating
quantities.
The Langevin equation for $g^i$ can then be written as
\eq
\frac{\partial g^{i}}{\partial t}=-\sum_{j}\nabla_{j}(
\sigma^{u}_{ij}+\sigma^{v}_{ij})-\sum_{j}\nabla_{j}(
\frac{g^i g^j}{\rho}
)-\sum_{j}L_{ij}(\frac{g^j}{\rho})+\Theta_{i},
\label{gi}
\en
where $L_{ij}({\bf x})=-\eta_{0}(\frac{1}{3}\nabla_{i}\nabla_{j}
+\delta_{ij}\nabla^{2})-\zeta_{0}\nabla_{i}\nabla_{j}$ with
the bare shear and bulk
viscosities, $\eta_{0}$ and
$\zeta_0$, respectively. For later use, we define the bare
longitudinal viscosity, $\Gamma_0 \equiv \zeta_0
+\frac{4}{3}\eta_0$. In Eq.~(\ref{gi}), the Gaussian noise
$\Theta_i$ satisfies
\eq
\langle\Theta_{i}({\bf x},t)\Theta_{j}
({\bf x}^{\prime},t^{\prime})\rangle =
2k_{B}TL_{ij}({\bf x})
\delta({\bf x-x^{\prime}})\delta(t-t^{\prime}).
\en
The streaming velocity for $n$ is given by
\eq
\bar{V}_n =\int d^3 {\bf x^\prime}\sum_i
\{ n({\bf x}),g^i({\bf
x^\prime})\}\frac{\delta F}{\delta g^i ({\bf x^\prime})}=
-\sum_i \nabla^i_{\bf x} \Bigl( n({\bf x})\frac{g^i ({\bf x})}
{\rho ({\bf x})}\Bigr) .
\en
Assuming $n$ is a diffusive variable, we must choose
$\Gamma_{nn}({\bf x-x^\prime})\equiv -\Gamma_v \nabla_{\bf x}^2
\delta ({\bf x-x^\prime})$ and the associated 
noise $\Xi$ satisfies
\eq
<\Xi({\bf x},t)\Xi({\bf x}^{\prime},t^{\prime})>=
  -2k_{B}T\Gamma_{v}\nabla^{2}_{x}\delta({\bf x-x^{\prime}})
  \delta(t-t^{\prime}) .
\en
We have the Langevin equation for $n$,
\eq
\frac{\partial n}{\partial t}=-\sum_{i}\nabla_{i}
(n\frac{g^{i}}{\rho})+\Gamma_{v}\nabla^{2}
\frac{\delta F}{\delta n}+\Xi .
\en
As in the case of $g^{i}$, we expand $n$ around the metastable
state $n^{*}[\delta\rho]$ 
and write the equation in terms of the fluctuation 
$\delta n$. Using Eq.~(\ref{expansion}) and keeping
up to quadratic terms in fluctuating variables, we have
\begin{eqnarray}
\frac{\partial}{\partial t}(\delta n)
&=& -\bar{n}(1+\frac{x}{\sigma y}){\bf
\nabla}\cdot ( \frac{{\bf g}}{\rho})
-{\bf\nabla}\cdot [
(\delta n)\frac{{\bf g}}{\rho}]  \label{n} \\
&&
+\bar{n}\widetilde{\Gamma}_v\nabla^2
\Bigl[ \sigma y(\frac{\delta n}{\bar{n}}
)-2x(\frac{1+\sigma}{\sigma})(\frac{\delta\rho}{\rho_0})
(\frac{\delta n}{\bar{n}})+y(1+\sigma )(\frac{\delta n}{
\bar{n}})^2\Bigr] +\Xi , \nonumber 
\end{eqnarray}
with the rescaled diffusive coefficient 
\eq
\widetilde{\Gamma}_v\equiv
\frac{A\rho^2_0}{\bar{n}^2}\Gamma_v .
\en

\subsection{Field-theoretic Formulation}

The Langevin equations, Eqs.~(\ref{rho}), (\ref{gi}) and
(\ref{n}) can be put into
a field theoretical form following 
standard Martin-Siggia-Rose (MSR)
procedures \cite{MSR,MSR1}. 
It essentially amounts to introducing the hatted
variable $\widehat{\psi}_\alpha$ for each field $\psi_\alpha$ to
enforce the equations of motion and integrating out the Gaussian
noises to get the quadratic form in $\widehat{\psi}_\alpha$.
We introduce the local velocity field
${\bf V}$, where ${\bf g}=\rho{\bf V}$,
to eliminate 
the $1/\rho$ nonlinearities \cite{dm}
in Eqs.~(\ref{gi}) and (\ref{n}).
This relation is then enforced by inserting the identity
\begin{eqnarray}
1 &=&\int\,{\cal D}{\bf V}\:\delta (\frac{{\bf g}}
{\rho}-{\bf V})=\|\rho\|\int\,{\cal D}{\bf V}\:
\delta ({\bf g}-\rho{\bf V}) \nonumber \\
&=&\|\rho\|\int\,{\cal D}{\bf \widehat{V}}
{\cal D}{\bf V}\:\exp \biggl( i\int\, d1\:
{\bf \widehat{V}}(1)\cdot \Bigl( {\bf g}(1)-\rho (1){\bf V}(1) 
\Bigr) \biggr)
\label{constraint}
\end{eqnarray}
into the functional integral form,
where $\|\rho\|$ is the functional Jacobian
between two delta functionals, and
$d1\equiv d^3 {\bf x}_1 dt_1 $, $ {\bf g}(1)\equiv
{\bf g}({\bf x}_1 ,t_1 )$, and so on.
As shown in Ref.~\onlinecite{my:1}, the Jacobian $\|\rho\|$ has
no dynamical effect, so that it can be neglected
throughout the analysis. 
The generating functional is then given in terms of a functional
integral over $\Psi_\alpha ,\widehat{\Psi}_\alpha$, where
$\Psi_\alpha =\{ \delta\rho , g_i, \delta n, V_i \}$. 
Without source terms,
it is given by
\eq
Z=\int\, {\cal D}\Psi{\cal D}\widehat{\Psi}
\:\exp(-S[\Psi ,\widehat{\Psi}]),
\en
where the action $S[\Psi ,\widehat{\Psi}]$ is given by
\widetext
\begin{eqnarray}
S[\Psi, \widehat{\Psi}]&=&\int\, d1\:
\biggl[ \:\sum_{ij}\widehat{g}_i \,\beta^{-1}
L_{ij}(1)\,\widehat{g}_j -\widehat{\delta n}\,
\beta^{-1}\Gamma_v \nabla^2
\,\widehat{\delta n} \label{S} \\
&&~~~~~~~+i\widehat{\delta\rho}\, \Bigl[ \frac{\partial 
\rho}{\partial t}
+{\bf \nabla}\cdot{\bf g}\, \Bigr] +i\sum_i\widehat{V}_i\, 
\Bigl[
g_i -\rho V_i \Bigr]  \nonumber \\
&&~~~~~~~+i\sum_i \widehat{g}_i \, \Bigl[
\frac{\partial g_i }{\partial t}
+\sum_j \nabla^j (\sigma^u_{ij} +\sigma^v_{ij})+\sum_j [
\nabla^j (\rho V_i V_j)+L_{ij}(1)(V_j )] \Bigr] \nonumber \\
&&~~~~~~~+i\widehat{\delta n}\, \Bigl[ \frac{\partial 
\delta n}{\partial t }+
\bar{n}(1+\frac{x}{\sigma y}){\bf
\nabla}\cdot {\bf V}
+{\bf \nabla}\cdot [ (\delta n){\bf V}] \nonumber \\
&&~~~~~~~~~~~~~~~~~-\bar{n}\widetilde{\Gamma}_v \nabla^2 [ \, 
\sigma y(\frac{\delta n}{\bar{n}})
-2x(\frac{1+\sigma}{\sigma})(\frac{\delta\rho }{\rho_0})
(\frac{\delta n}{\bar{n}})+(1+\sigma )y(
\frac{\delta n}{\bar{n}})^2\, ] \: \Bigr] \biggr] , \nonumber 
\end{eqnarray}
\narrowtext
where $\beta^{-1}=k_B T$.
As a result of Eq.~(\ref{constraint}), we have a polynomial 
action in $\Psi_\alpha$ and $\widehat{\Psi}_\alpha$, to which
we can apply the perturbation theory expansion in a standard way.
The nonlinear corrections coming from the cubic 
terms \cite{fn:3} in the action generate the one-loop 
self-energies $\Sigma_{\alpha
\beta}$ which modify the zeroth order inverse propagators
$[G^0_{\alpha\beta}]^{-1}$ through Dyson's equation:
\eq
G_{\alpha\beta}^{-1}=[G^0_{\alpha\beta}]^{-1}-\Sigma_
{\alpha\beta}.
\label{Dyson}
\en

\subsection{Linearized Theory}

From the quadratic terms in the action, Eq.~(\ref{S}), we
can easily read off the elements of the zeroth order inverse
propagator, $[G^0_{\alpha\beta}]^{-1} $. 
By inverting this matrix, we get various 
correlation and linear response functions. For later use,
we list here a few of these functions.
In this paper, we will be concerned only with
the longitudinal parts of those functions
and use
hereafter the notation
$\rho , n$ instead of $\delta\rho , \delta n$ respectively
for convenience. In the Fourier-transformed
space, we have
\begin{eqnarray}
&&G^{0}_{n\widehat{n}}({\bf q},\omega )
=-G^{0\, *}_{\widehat{n}n}({\bf q},\omega )
=\frac{D_0}{W_0}, 
\label{Gnhatn} \\
&&G^0_{\rho\widehat{\rho}}({\bf q},\omega )=
\frac{1}{W_0}\Bigl[
(\omega +i\sigma y\widetilde{\Gamma}_v q^2 )(\rho_0
\omega +i\Gamma_0 q^2 )-A\rho^2_0
\frac{(x+\sigma y)^2}{\sigma y}q^2 \Bigr] ,
\label{Grhatr}
\end{eqnarray}
where
\begin{eqnarray}
&&D_{0}\QO =\rho_{0}(\omega^{2}-q^{2}c^{2}_{0})+i\omega q^{2}
\Gamma_{0}, \label{D0} \\
&&W_{0}\QO =(\omega +i\sigma y\widetilde{\Gamma}_v
q^{2})D_{0}-\frac{(x+\sigma y)^2}
{\sigma y}\omega q^{2} \label{W0} .
\end{eqnarray}
The density-density and defect-defect correlation
functions are given respectively by
\begin{eqnarray}
&&G^{0}_{nn}({\bf q},\omega )=\bar{n}^2\frac{2\beta^{-1}q^{2}}
{|W_{0}|^{2}}\Bigl[ (\frac{x+\sigma y}{\sigma y})^2
\Gamma_{0}\omega^{2}q^{2}
+\frac{\widetilde{\Gamma}_v}{A\rho^2_0}|D_{0}|^{2}\Bigr] , 
\label{Gnn} \\
&&G^{0}_{\rho\rho}({\bf q},\omega )=\rho^2_0
\frac{2\beta^{-1}
q^{4}}{|W_{0}|^{2}}\Bigl[ \Gamma_{0}|\omega +i\sigma y
\widetilde{\Gamma}_{v}q^{2}|^{2}
+A\rho^2_0 (x+\sigma y)^2
\widetilde{\Gamma}_{v}q^{2}\Bigr] \label{Grr} .
\end{eqnarray}
Other correlation functions involving $n$ are given by
\eq
G^{0}_{\rho n}({\bf q},\omega )=\rho_0\bar{n}
\frac{2\beta^{-1}
q^{4}}{|W_{0}|^{2}}(\frac{x+\sigma y}{\sigma y})
\Bigl[\omega^{2}\Gamma_{0}
+\sigma y\widetilde{\Gamma}_{v}\rho_0(\omega^{2}-q^{2}c^{2}_{0})
\Bigr] , 
\label{Grn} 
\en
and
\eq
G^{0}_{gn}({\bf q},\omega )=\rho_{0}
G^{0}_{Vn}({\bf q},\omega )=(\frac{\omega}{q})
G^{0}_{\rho n}({\bf q},\omega ) \label{misc}.
\en
We note that the zeroth order response and correlation functions
satisfy the following fluctuation-dissipation theorem:
\begin{eqnarray}
&&G^0_{\alpha n}({\bf q}, \omega )=-
\frac{1}{\sigma y}(\frac{\bar{n}}{\rho_0})^2
\frac{2\beta^{-1}}{A}
{\rm Im}\: G^0_{\alpha\widehat{n}}({\bf q},\omega ), \nonumber \\
&&G^0_{\rho\rho}({\bf q},\omega )=-\frac{2\beta^{-1}
}{A}{\rm Im}\: G^0_{\rho\widehat{\rho}}({\bf q},\omega ),
\end{eqnarray}    
where $\alpha =\{\rho , g_i , V_i, n\}$.

From Eqs.~(\ref{Gnhatn}-\ref{W0}), we find that if
we can neglect the factor
\eq
\left|\frac{(x+\sigma y)^2}{\sigma y}\right|=
\left|\sigma y(1-a_1 )^2\right|\ll 1,
\label{s00}
\en
then we have
\eq
G^0_{n\widehat{n}}\QO\simeq\frac{1}{\omega +i\gamma_v}
\label{corn:lin} 
\en
and
\eq
G^0_{\rho\widehat{\rho}}\QO\simeq\frac{\rho_0 \omega+
i\Gamma_0 q^2}{D_0 \QO },
\label{corr:lin}
\en
where $\gamma_v\equiv \sigma y\widetilde{\Gamma}_v q^2$.
The assumption in Eq.~(\ref{s00}) is equivalent to
saying that we are considering the small $|\sigma |$ limit,
which will be justified self-consistently later
when we consider the conditions for the slowing down.
Eq.~(\ref{corr:lin}) represents a standard form for the density
auto-correlation function. The defect auto-correlation function
in Eq.~(\ref{corn:lin}) 
has a very slow diffusive mode, $\omega =-i\gamma_v$, since 
it is assumed that
$\gamma_v\ll 1$. This signals a separation
of time scales between the density and the defect variables as
noted in Sec.~\ref{sec:intro}. 

We can achieve a further simplification by looking at the following
situation.
As one approaches the glass
transition,
we expect that, since the viscosity is getting extremely large,
one eventually 
reaches a point where
\eq
\frac{q^{2}\Gamma_{0}}{\rho_{0}}\gg\omega .
\en
Then, from Eqs.~(\ref{W0},\ref{s00}), the denominator $W_{0}$ in
Eqs.~(\ref{Gnhatn}) and (\ref{Grhatr}) reduces to
\begin{eqnarray}
W_{0}&\simeq & iq^{2}
\Gamma_{0}\{\omega^{2}+i(\gamma_v
+\gamma_u)\omega-\gamma_{v}\gamma_{u}\} \nonumber \\
&=& iq^2 \Gamma_0 (\omega +i\gamma_v )(\omega +i\gamma_u ),
\end{eqnarray}
where $\gamma_u\equiv A\rho^2_0 /\Gamma_0$.
It follows then from Eqs.~(\ref{Gnn}-\ref{Grn}) that
\begin{eqnarray}
G^{0}_{nn}({\bf q},\omega )&\simeq& 
\frac{1}{\sigma y}(\frac{\bar{n}}{\rho_0})^2\,\frac{2\beta^{-1}}
{A}\;
\frac{\gamma_{v}}{\omega^{2}+\gamma^{2}_{v}},  
\label{Gnn:app} \\
G^{0}_{\rho\rho}({\bf q},\omega )&\simeq  &\, 
\frac{2\beta^{-1}}
{A}\;\frac{\gamma_{u}}{\omega^{2}+\gamma^{2}_{u}},
\label{Grr:app} \\
G^{0}_{\rho n}({\bf q},\omega )&\simeq &\frac{\bar{n}}{\rho_0}\,
(\frac{x+\sigma y}{\sigma y})\,\frac{2\beta^{-1}}{A}
\;(-\frac{\gamma_{v}}{\omega^{2}+\gamma^{2}_{v}}
+\frac{\gamma_{u}}{\omega^{2}+\gamma^{2}_{u}})
\nonumber \\
&\simeq & -\frac{\rho_0}{\bar{n}}\,
(x+\sigma y)\, G^{0}_{nn}({\bf q},\omega )+\frac{\bar{n}}
{\rho_{0}}\,(\frac{x+\sigma y}{\sigma y})\,
G^{0}_{\rho\rho}({\bf q},\omega ).
\label{Grn:app}
\end{eqnarray}
Therefore, in this approximation,
we have the density-defect correlation function
as a linear combination of the density and the defect
auto-correlation functions.

\subsection{ Nonlinear
Corrections}

Nonlinear corrections to the zeroth order
response and correlation functions in 
the previous section are represented in terms of 
the self-energies through Eq.~(\ref{Dyson}). For
example, Eqs.~(\ref{Grhatr},\ref{Gnhatn})
are modified to yield very complicated
expressions:
\begin{eqnarray}
G_{n\widehat{n}}\QO &=& \frac{D\QO }{W\QO } \label{G_nhn} \\
G_{\rho\widehat{\rho}}({\bf q},\omega )&=&\frac{1}
{W({\bf q},\omega )}\Bigl[ \: [\omega +i\gamma^\prime_v
({\bf q},\omega ) ][\rho({\bf q},\omega )\omega
+i\Gamma ({\bf q},\omega )q^2 ]  \nonumber \\
&&~~~~~~~~~~-q \bar{n}^\prime ({\bf q},\omega )
[qu^\prime({\bf q},\omega )+i\omega
\Sigma_{\widehat{V}n}({\bf q},\omega ) ]\;\Bigr] ,
\label{G_rhr} 
\end{eqnarray}
where
\begin{eqnarray}
W({\bf q},\omega )&\equiv &[\omega +i\gamma^\prime_v
({\bf q},\omega ) ] D({\bf q},\omega ) \label{W} \\
&&-q^2 u^\prime ({\bf q},\omega ) \Bigl[
\bar{n}^\prime ({\bf q},\omega )[\omega
+iq\Sigma_{\widehat{V}\rho}({\bf q},\omega ) ]
+\rho ({\bf q},\omega )\Sigma_
{\widehat{n}\rho}({\bf q},\omega )\;\Bigr] \nonumber \\
&&-iq\Sigma_{\widehat{V}n}({\bf q},\omega )\Bigl[
\bar{n}^\prime ({\bf q},\omega )[\omega^2
-q^2 c^2 ({\bf q},\omega )]-i\Gamma ({\bf q},\omega )
q^2 \Sigma_{\widehat{n}\rho}({\bf q},\omega )]\;\Bigr],
\nonumber
\end{eqnarray}
and 
\eq
D({\bf q},\omega )\equiv \rho ({\bf q},\omega )
[\omega^2 -q^2 c^2 ({\bf q},\omega )]+i\Gamma
({\bf q},\omega )q^2 [\omega +iq\Sigma_{\widehat{V}\rho}
({\bf q},\omega )], \label{D}
\en
with the renormalization of parameters given by
\begin{mathletters}
\begin{eqnarray}
&&\rho ({\bf q},\omega )=\rho_0 -i\Sigma^L_{\widehat{V}V}
({\bf q},\omega ) \\
&&qc^2 ({\bf q},\omega )=qc^2_0 +\Sigma_{\widehat{g}\rho}
({\bf q},\omega ) \\
&&q^2\Gamma ({\bf q},\omega )=q^2\Gamma_0 +i
\Sigma^L_{\widehat{g}V}({\bf q},\omega ) \label{Gammaqo} \\
&&\gamma^\prime_v ({\bf q},\omega ) =
\gamma_v
+i\Sigma_{\widehat{n}n}({\bf q},\omega ) \label{Gv:ren} \\
&&q\bar{n}^\prime ({\bf q},\omega )=q(1-a_1 )\bar{n}
+\Sigma_{\widehat{n}V}({\bf q},\omega ) \\
&&qu^\prime ({\bf q},\omega )=q(B\rho_0 +
\sigma\epsilon\bar{n}^3 )
+\Sigma_{\widehat{g}n}({\bf q},\omega ) \nonumber \\
&&~~~~~~~~~~~~~=(x+\sigma y)\frac{A\rho^2_0}{\bar{n}}
+\Sigma_{\widehat{g}n}({\bf q},\omega ). \label{u:ren}
\end{eqnarray}
\end{mathletters}
In the above equation $L$
denotes the longitudinal part of self-energy. For example,
$\Sigma_{\widehat{V}_i V_j}\QO$ is decomposed into the longitudinal
and transverse parts:
\eq
\Sigma_{\widehat{V}_i V_j}\QO =\frac{q_i q_j}{q^2}
\Sigma^L_{\widehat{V}V}\QO
+(\delta_{ij}-\frac{q_i q_j}{q^2} )\Sigma^T_{\widehat{V}V}\QO.
\en
As shown by Das and Mazenko 
\cite{dm}, the self-energy $\Sigma_{\widehat{V}\rho}\QO $
is responsible for the cutoff mechanism. In addition to
that, we have self-energies $\Sigma_{\widehat{V}n}\QO$
and $\Sigma_{\widehat{n}\rho}\QO$ in this case. As will 
be discussed
in Sec.~\ref{sec:cut}, all three self-energies are related to 
the cutoff mechanism. Although the
final stage of relaxation is governed by this
cutoff effect, we neglect it in the present analysis. 

The separation of time scales observed in the zeroth order
correlation functions remains valid at high orders if the renormalized
version of Eq.~(\ref{s00}) holds, {\it i.e.},
\eq
\left| \frac{u^\prime \QO \bar{n}^\prime \QO}
{\rho\QO c^2\QO} \right|\:\ll 1.
\en
In this case, Eqs.~(\ref{G_nhn},\ref{G_rhr}) give us 
\begin{eqnarray}
&&G_{n\widehat{n}}\QO\simeq\frac{1}{\omega +i\gamma^\prime_
v \QO} \label{original:n} \\
&&G_{\rho\widehat{\rho}}({\bf q},\omega )\simeq 
\frac{\rho({\bf q},\omega )\omega
+i\Gamma ({\bf q},\omega )q^2}{
\rho ({\bf q},\omega )
[\omega^2 -q^2 c^2 ({\bf q},\omega )]+i\Gamma
({\bf q},\omega )\omega q^2 
}.
\label{original:r}
\end{eqnarray}
Eqs.~(\ref{original:n}) and (\ref{original:r}) are the 
fundamental equations for the dynamics of density 
and defect fluctuations without the cutoff effect.
In the following, we calculate explicitly nonlinear
contributions to the renormalized 
viscosity $\Gamma\QO$ and the renormalized diffusion
coefficient $\gamma^\prime_v\QO$, which will
be expressed back in terms of $G_{n\widehat{n}}\QO$ and
$G_{\rho\widehat{\rho}}({\bf q},\omega )$.

\subsection{One-Loop Evaluation of $\Gamma\QO$ and 
$\gamma^\prime_v\QO$}
The density feedback mechanism of MCT is realized by
calculating the nonlinear corrections to the bare
viscosity in the density correlation 
function of the form,
Eq.~(\ref{original:r}).
In the present case, however, we have two coupled equations,
(\ref{original:n}) and (\ref{original:r}). Thus we must find
both kinetic
coefficients, $\Gamma\QO$ and $\gamma^\prime_v\QO$ to
complete the equations. 
In this section, we calculate one-loop 
nonlinear contributions
to those quantities. 

\vskip8mm
\epsfxsize=8cm
\centerline{\epsfbox{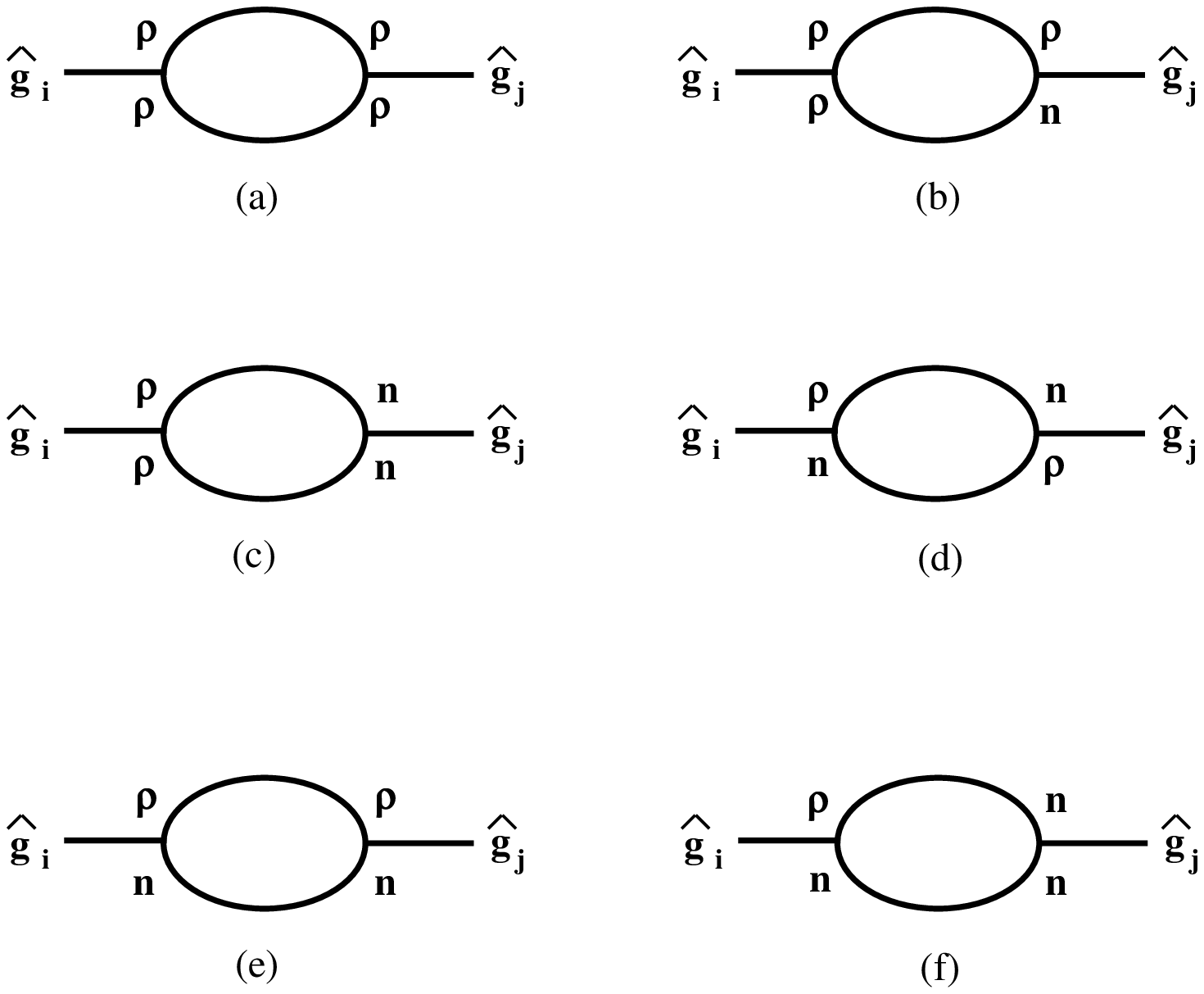}}
\centerline{\small One-loop diagrams contributing to $\Sigma_
{\widehat{g}\widehat{g}}\QO$ }
\centerline{\sc Fig.~2}

As noted in Ref.~\onlinecite{dm}, the renormalized longitudinal
viscosity in Eq.~(\ref{Gammaqo}) can also be represented in the
hydrodynamic limit as
\eq
\Gamma ({\bf q},\omega )=\Gamma_{0}-\frac{\beta}{2q^2}
\Sigma^{L}_{\widehat{g}\widehat{g}}({\bf q},\omega ).
\label{visc:long}
\en
The relevant
one-loop diagrams contributing to this self-energy
are listed in Fig.~2. As discussed in Ref.~\onlinecite{dm},
the diagrams coming from
the convective vertex ($\widehat{g}\rho V V$ vertex) in the action
Eq.~(\ref{S}) just renormalizes the bare viscosity and
are irrelevant to the density feedback mechanism.

We note that the approximation given by Eq.~(\ref{Grn:app})
generates two kinds of terms that are proportional to
$G_{nn}G_{\rho\rho}$ and $G_{\rho\rho}G_{\rho\rho}$,
respectively,
in the expression for $\Gamma({\bf q},\omega )$. 
In fact, one can easily see that the 
$G_{nn}G_{\rho\rho}$ type terms come from
diagrams (b),(c),(d),(e) and (f), and the 
$G_{\rho\rho}G_{\rho\rho}$ type terms from
diagrams (a),(b),(c) and (d) in Fig.~2. 
Thus we have 
\widetext
\begin{eqnarray}
\Gamma ({\bf q},\omega )&=&\Gamma_0 +\int^\infty_0
dt\;  e^{i\omega t}\int\frac{d^3 {\bf k}}{(2\pi )^3 } 
\label{G1} \\
&&~~~~~~~~~~~~~~
\biggl[ \Bigl[ V^{(1)} ({\bf q},{\bf k})G_{nn}({\bf k},t)
+V^{(2)}({\bf q},{\bf k})
G_{\rho\rho}({\bf k},t)\Bigr] G_{\rho\rho}({\bf q-k},t)
\biggr] , \nonumber
\end{eqnarray}
\narrowtext
where $V^{(1)}$ and $V^{(2)}$ are appropriate vertices 
to be evaluated.
In principle, the wave number dependence of the vertices can be
considered\cite{Das}, for example, 
by using the spatial derivatives of
the density fluctuations in the effective Hamiltonian,
Eq.~(\ref{Fu}). This involves, however, very complicated wavenumber
integrals. In this analysis, we consider the 
wavenumber independent case. A closely related 
approximation to this is that the correlation functions can be
factorized into wavenumber and time dependent parts 
\cite{factorization}. We assume here that
$G_{nn} ({\bf q},t)=T(q)\psi (t)$,
$G_{\rho\rho}({\bf q},t)=S(q)\phi (t)$,
$T(q)$ and $S(q)$ are the flat structure factors given by
\eq
T(q)=(\frac{\bar{n}}{\rho_0})^2\frac{1}{\sigma y}\frac{\beta^{-1}}
{A},~~~~~~~S(q)=\frac{\beta^{-1}}{A},
\label{flat}
\en
for $q<\Lambda$ and $T(q)=S(q)=0$ for $q>\Lambda$, where
$\Lambda$ is the large momentum cutoff.
Integrating over wavenumbers
for the one-loop diagrams in Fig.~2 with the help of 
$\Lambda$, we obtain 
\eq
\Gamma (\omega)=  \Gamma_0 +k_B T\frac{\Lambda^3}{6\pi^2}
\int^\infty_0 \; dt\: e^{i\omega t}[d_1 \psi (t)\phi (t)
+d_2 \phi^2(t)],
\label{Gamma(t)}
\en
where the coefficients $d_1$ and $d_2$ 
are functions of $x,y$ and $\sigma$.
We find, through detailed calculation of the diagrams
that
\begin{mathletters}
\label{d12:0}
\begin{eqnarray}
d_1 (x,y,\sigma )&=&
-4x(\frac{1+\sigma}{\sigma})^2 (2+\frac{x}{\sigma y}) 
\label{l1} \\
&&-4\Bigl[ y(2+3\sigma )+2x^2 (\frac{1+\sigma }{\sigma })^2 
\Bigr]
\, (1+\frac{x}{\sigma y})^2 ,\nonumber \\
d_2 (x,y,\sigma )&=&1-
4x(\frac{1+\sigma }{\sigma})(1+\frac{x}{\sigma y}) 
\label{l2} \\
&&+2\Bigl[ y(2+3\sigma )+2x^2 (\frac{1+\sigma }{\sigma })^2 
\Bigr]
\, (1+\frac{x}{\sigma y})^2 .\nonumber
\end{eqnarray}
\end{mathletters}
For later use, we define a 
dimensionless parameter containing an explicit factor of the
temperature, 
\eq
\xi\equiv\frac{k_B T}{A\rho^2_0}\,\frac{\Lambda^3}{6\pi^2}.
\en

The renormalization of $\gamma_v$ is given by
Eq.~(\ref{Gv:ren}). As in Eq.~(\ref{visc:long}),  
it is equivalent in the hydrodynamic
limit to
\eq
\gamma^\prime _v ({\bf q},\omega )=
\gamma_v
-\case{1}{2}\sigma y(\frac{\rho_0}{\bar{n}})^2
\frac{A}{\beta^{-1}}\Sigma_
{\widehat{n}\widehat{n}}({\bf q},\omega ).
\en
Among the nonlinear terms in Eq.~(\ref{S}) that contribute,
at the one-loop order,
to the self-energy $\Sigma_{\widehat{n}\widehat{n}}({\bf q},
\omega)$, those of $\widehat{n}\rho n$ and 
$\widehat{n} n n$ types contain an explicit 
factor of $\Gamma_v$ which is assumed to be very small. 
The other one-loop diagrams that contribute
to $\Sigma_{\widehat{n}\widehat{n}}({\bf q},\omega )$,
and do not have the
explicit $\Gamma_v$ factor are listed in Fig.~3.

\vskip8mm
\epsfxsize=8cm
\centerline{\epsfbox{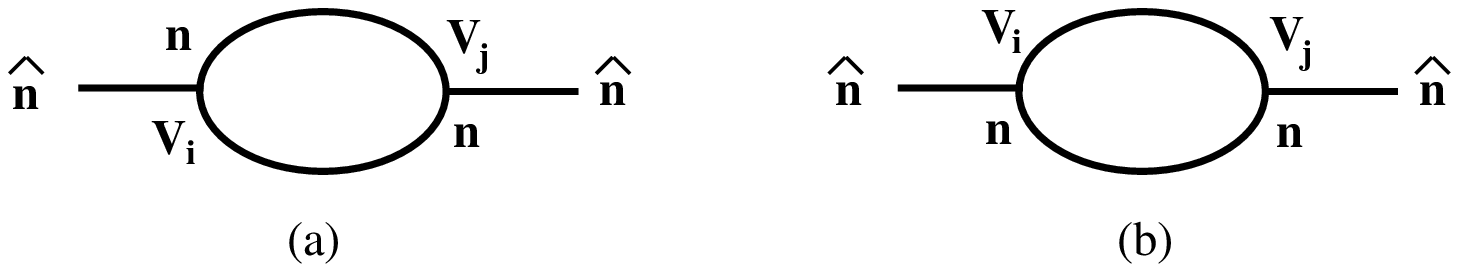}}
\centerline{\small One-loop diagrams contributing to
$\Sigma_{\widehat{n}\widehat{n}}\QO$}
\centerline{\sc Fig.~3}

Evaluating the diagram (a) using Eqs.~(\ref{misc}) and 
(\ref{Grn:app}), we find that 
this contribution is directly proportional to
\eq
\left| \frac{(x+\sigma y)^2}{\sigma y}\right|\ll 1,
\en
which is our self-consistent assumption, Eq.~(\ref{s00}).
The evaluation of the second diagram (b) yields
\eq
\case{1}{2}\sigma y(\frac{\rho_0}{\bar{n}})^2
\frac{A}{\beta^{-1}}\Sigma^{(b)}_
{\widehat{n}\widehat{n}}({\bf q},\omega )
=(\frac{q}{\Lambda})^2\xi\;\int^\infty_0
dt\; e^{i\omega t}\:\psi (t)\frac{d^2\phi (t)}
{dt^2}\: +\: I_T ,
\label{Snn:b}
\en
where we have used Eq.~(\ref{misc}).
The second term $I_T$ in Eq.~(\ref{Snn:b}) is the integral
containing the transverse part of the velocity auto-correlation
function, $G^T_{VV}\QO$. We note that in the framework of the 
flat structure factor, Eq.~(\ref{flat}), there is no
coupling between the density feedback mechanism and the transverse 
viscosity. Thus, we assume that this term is insensitive to any
dramatic slowing down.
We obtain finally that
\eq
\gamma^\prime_v\QO=\gamma_v
-(\frac{q}{\Lambda})^2\xi\;\int^\infty_0
dt\; e^{i\omega t}\:\psi (t)\ddot{\phi}(t).
\label{new}
\en
Eqs.~(\ref{original:n}) and (\ref{original:r}) together with
Eqs.~(\ref{Gamma(t)}) and (\ref{new}) complete the specification
of our model. 

\section{The model with a constant
defect auto-correlation function}
\label{sec:cnn1}
Metastability plays an important role in this model, since
defects spend considerable amount of time trapped within the 
well. During this period, we may assume that the defect auto-correlation
function can be regarded as a constant:
\eq
G_{nn}\QO=2\pi (\frac{\bar{n}}{\rho_0})^2\frac{1}{\sigma y}\frac{
\beta^{-1}}{A}\;\delta (\omega ),~~~~~{\rm or}~~~~~\psi (t)=1.
\label{psi1}
\en
This approxiamtion corresponds to the limit where one
takes the bare value of
$\gamma_v$ to zero. In this case, we note that the nonlinear 
correction to $\gamma_v$ also vanishes in the long time
limit, $\omega\rightarrow 0$, since the integrand in
Eq.~(\ref{new}) is a total derivative when $\psi (t)=1$. Thus
the self-consistency of the approximation is maintained.

The coupled equations (\ref{original:n}) and (\ref{original:r})
then reduce to a form dependent only on $\phi (t)$ and
we can make contact with the standard treatment of MCT
\cite{mct:gen}. 
Indeed Eqs.~(\ref{original:r}) and (\ref{Gamma(t)}) 
with (\ref{d12:0}) are just
$N=2$ case of the standard MCT equations, (\ref{1}) and (\ref{Gamma}),
where only the linear $(c_1)$ and quadratic $(c_2)$ terms
are considered. We can identify $
\Omega^2_0 =q^2 c^2$, $d_0 =q^2 \Gamma_0 /\rho$,
and
\eq
c_1 =\xi d_1 (x,y,\sigma ),~~~~~c_2 =\xi d_2 (x,y,\sigma ).
\label{cd12}
\en
It is important to note that the MCT coefficients $c_1$ and
$c_2$ are expressed in terms of the temperature
and the parameters describing
the metastable potential and the coupling between the density
and the defect variables.
We now assume that the system organizes itself to be on the
critical surface of MCT, which is described below. This
will give relations among the metastability parameters.

The density feedback mechanism associated with the
representation of the type 
Eq.~(\ref{1}) with $N=2$ was first
studied by G\"otze \cite{Gotze:lin}. According to
Ref.~\onlinecite{Gotze:lin}, there exists a critical line
in the $( c_1 , c_2 )$ space
separating ergodic and nonergodic regions, where
the nonergodic phase is characterized by
the existence of the limit $f=
\lim_{t\rightarrow\infty}\phi (t),~~f>0$. 
To study the critical condition in terms of the
parameters $x,y,\sigma$ and $\xi$,
we follow a more general discussion on such model
given
by Kim and Mazenko \cite{km}. According to
Ref.~\onlinecite{km}, the glass transition
can be described by the following three
parameters:
\begin{eqnarray}
&&\sigma_0 =(1-f)V(f) \label{s0}\\
&&\sigma_1 =(1-f)^2 V^\prime (f) \label{s1} \\
&&\lambda =\case{1}{2}(1-f)^3 H^{\prime\prime}(f), \label{lf}
\end{eqnarray}
where $V(f)=H(f)-f/(1-f)$ and $H(f)=\sum^N_{i=1}c_i f^i$
for a general model containing higher order terms
in the mode coupling integral.
In Eq.~(\ref{lf}), the parameter $\lambda$ is directly related 
to the exponents of the sequence of time relaxations 
\cite{Gotze:lin}:
\eq
\frac{\Gamma^2 (1-a)}{\Gamma (1-2a)}=\lambda =
\frac{\Gamma^2 (1+b)}{\Gamma (1+2b)}.
\label{ablam}
\en
The ideal glass transition is approached when both $\sigma_0$
and $\sigma_1$ are getting small. Let us consider
the situation where $\sigma_1 =0$ and the transition
is approached by taking $\sigma_0 \rightarrow 0$.
Solving Eqs.~(\ref{s0}-\ref{lf}) for $c_1$ and
$c_2$, we have
\begin{mathletters}
\label{c1212}
\begin{eqnarray}
&&c_1 =\frac{2\lambda -1}{\lambda^2}+\frac{4\sigma_0}
{\lambda (1-\lambda )}+{\cal O}(\sigma^2_0 ) \label{c1s0} \\
&&c_2 =\frac{1}{\lambda^2}-\frac{3\sigma_0}
{\lambda (1-\lambda )^2}+{\cal O}(\sigma^2_0 ). \label{c2s0}
\end{eqnarray}
\end{mathletters}
In terms of the parameters,
$x, y ,\sigma$ and $\xi$, the critical condition
is given by setting $\sigma_0 =0$:
\begin{mathletters}
\label{d12:cr}
\begin{eqnarray}   
&& d_1 (x, y ,\sigma )=\frac{2\lambda -1}{\xi\lambda^2}, 
\label{crit:f1} \\
&& d_2 (x, y ,\sigma )=\frac{1}{\xi\lambda^2},~~~~~~~\frac{1}{2}
<\lambda <1. \label{crit:f2}
\end{eqnarray}
\end{mathletters}
We note that we are mainly concerned with the region
where $|x|\ll 1$ in order to be consistent with the fact that we 
considered the coupling term as a small perturbation
shifting the metastable state as in Eqs.~(\ref{expansion})
and (\ref{abll1}).
We also require the solution be consistent with
the assumption, Eq.~(\ref{s00}).
We find
that nontrivial solutions 
to Eq.~(\ref{d12:cr}) 
satisfying Eqs.~(\ref{abll1})
and (\ref{s00}) exist only if we take 
$\sigma\rightarrow 0$ simultaneously with $x
\rightarrow 0$, while holding
\eq
\frac{x}{\sigma^2}\rightarrow C ,\label{limit}
\en
for some constant $C$.
Let us briefly discuss the physical meaning of Eq.~(\ref{limit}).
The small $|\sigma |$ limit is a sufficient
condition for a separation of the time scales between
the defect and the density variables as seen from Eq.~(\ref{s00}).
This indicates that the metastable wells become very shallow.
If the coupling energy represented by $|x|$ is stronger 
than the above limit, {\it i.e.} $|x| \sim {\cal O}(|\sigma |)$,
then the metastable state given by Eq.~(\ref{expansion})
does not exist, since the condition Eq.~(\ref{abll1})
is violated. Thus, a strong coupling energy destroys the 
metastability of defects. On the other hand, if 
$|x|\sim {\cal O}(|\sigma |^3 )$, we have from Eq.~(\ref{d12:0}),
\eq
d_1 =-8y,~~~~~d_2=1+4y .
\en
Since $y$ is always positive, $d_1 <0$, therefore
the system never reaches the critical surface given by
Eq.~(\ref{d12:cr}). Thus,
if the coupling is weaker than the limit given by Eq.~(\ref{limit}),
the slow dynamics of defects can not affect
the density dynamics so that the density variable does
not slow down. Eq.~(\ref{limit}) gives the correct 
relation between the coupling energy and the barrier size 
of the defects in the case of observed slowing down.
In this limit, Eq.~(\ref{d12:0}) reduce to 
\begin{mathletters}
\label{d12:re}
\begin{eqnarray}
&&d_1 (x,y,\sigma )= -8(y+C)-4(\frac{C^2}{y}+3y+8C)\sigma
+{\cal O}(\sigma^2 ), \label{f1s} \\
&&d_2 (x,y,\sigma )= 1+4y +2(3y+2C)\sigma
+{\cal O}(\sigma^2 ). \label{f2s}
\end{eqnarray}
\end{mathletters}
Thus, from Eqs.~(\ref{d12:cr}) and
(\ref{d12:re}), we have the 
critical condition given in terms of the metastability
parameters by
\eq
C=\frac{1}{4}(1-\frac{2\lambda+1}{2\xi\lambda^2}),~~~~~
y=\frac{1}{4}(\frac{1}{\xi\lambda^2}-1).
\label{Cy:crit}
\en
Furthermore, from Eqs.~(\ref{c1212}) and 
(\ref{d12:re}), we can easily see that the limit
$\sigma\rightarrow 0$ can be identified with
$\sigma_0\rightarrow 0$ if
\eq
3(\frac{C^2}{y}+3y+8C)=2(1-\lambda )(3y+2C).
\label{C*y}
\en
Therefore, as indicated in Sec.~\ref{sec:intro}, we have
the situation where
the stretched dynamics is associated with the weak
coupling $(x\rightarrow 0)$ and low barrier 
$(\sigma\rightarrow 0)$ limit, which
is consistent with the separation of the time scales between
the density and the defect variables as can be seen from
Eq.~(\ref{s00}). The condition Eq.~(\ref{C*y}),
when we use Eq.~(\ref{Cy:crit}), can be 
interpreted as a relation between the exponent parameter
$\lambda$ and the temperature represented by the parameter $\xi$: 
\eq
\xi=\frac{1}{\lambda^2}\Bigl[ 1-\frac{3(2\lambda -1)}{
2[7+2\lambda +\sqrt{4\lambda^2 +22\lambda +91}] }\Bigr] .
\label{lxi}
\en
Thus the exponents $a$ and $b$ are given as smooth functions 
of temperature. (See Fig.~4.) The temperature dependence of the
parameters $y$ and $C$ which describe the potential $h(n)$ and
its coupling to $\rho$ is given by Eq.~(\ref{Cy:crit}). 
All the temperature dependences certainly 
do not show any indication of a special temperature.
In the conventional MCT \cite{leut,Gotze:1}, the control parameter
$\sigma_0$ is assumed to have a temperature dependence as
$\sigma_0\sim T_0 -T$, which is the origin of the sharp
temperature dependences. In this case, however, $\sigma_0$
is proportional to $\sigma$ with weakly temperature-dependent
coefficient $f(\xi )$:
\eq
\sigma_0=f(\xi )\sigma\equiv
-\frac{2}{3}[3y+2C]\xi\lambda(1-\lambda )^2\;
\sigma .
\label{sigsig0}
\en
We note that $f(\xi )$ vanishes at the lower and upper bounds of $\xi$
which correspond to $\lambda =1$ and $\frac{1}{2}$,
respectively. Therefore, the
above analysis is not applicable to the region near the
two end points of $\xi$.
Our basic picture of the observed slowing down
of the dynamics is that
the parameters describing the metastable wells and the coupling,
while displaying a smooth temperature dependence,
achieve the critical limit given by Eq.~(\ref{limit}).
In this model, the transition is actually
controlled by the parameter $\sigma$, which is
the size of barrier in the metastable potential.

\vskip8mm
\epsfxsize=8cm
\centerline{\epsfbox{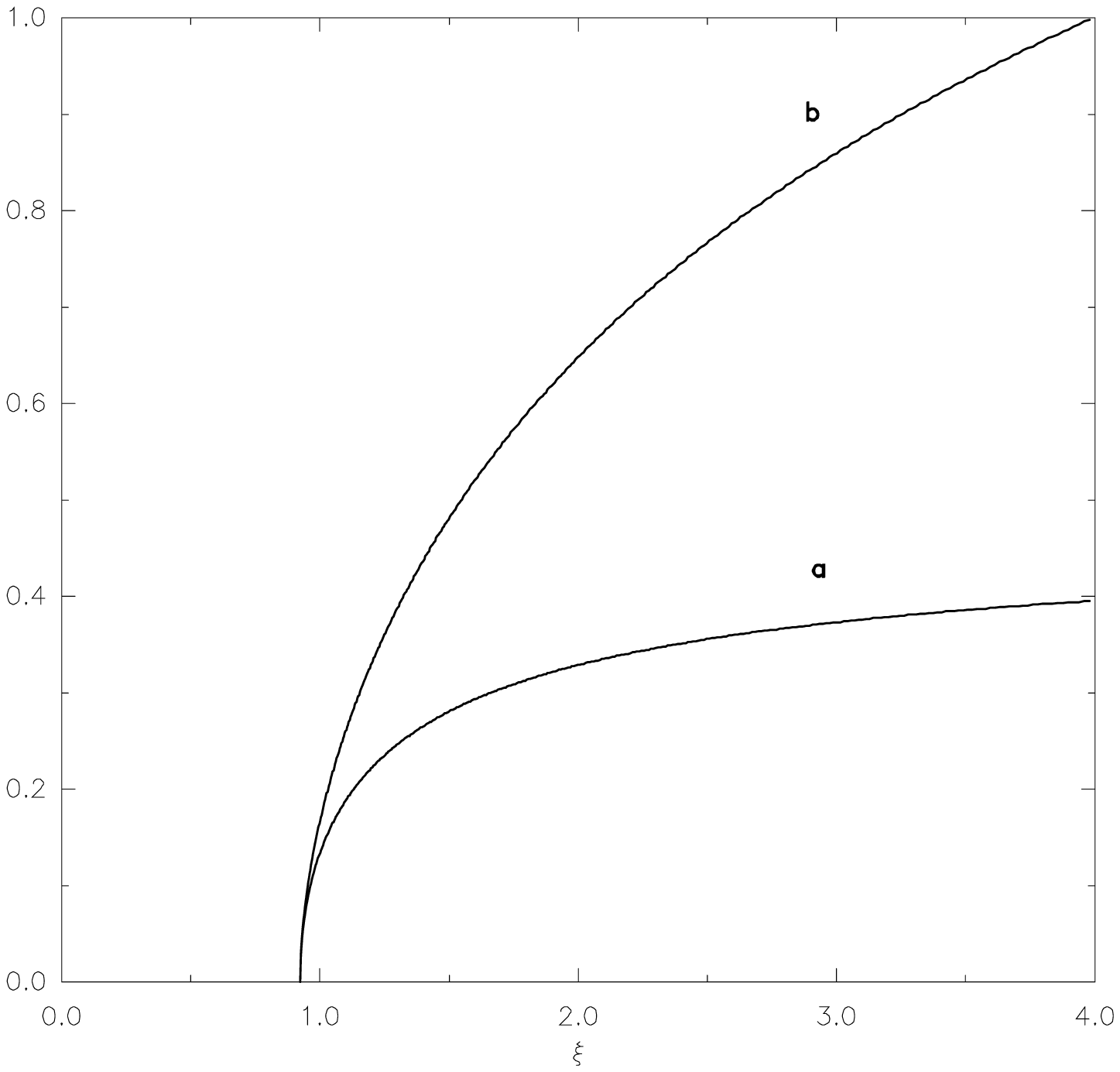}}
\centerline{\small The exponents $a$ and $b$ as functions of temperature
represented by $\xi$}
\centerline{\sc Fig.~4}

\section{The model including the dynamics of defects}
\label{sec:main}

The approximation, Eq.~(\ref{psi1}), made in the previous
section enabled us to treat the model analytically and to
obtain the critical condition between the shallow
metastable wells and the weak coupling.
However, when nonlinear corrections to the bare diffusion
coefficient come into play,
the constraint
Eq.~(\ref{psi1}) must be relaxed and the dynamics
of defects should be taken into account. 
In this section, we analyze the full coupled equations, (\ref{original:n})
and (\ref{original:r}) with (\ref{Gamma(t)}) and (\ref{new}).
Since the nonlinear couplings in these equations are highly nontrivial, 
an analytical treatment is very difficult.
In the time domain,
these equations are coupled integro-differential equations:
\begin{eqnarray}
&&\ddot{\phi}(t)+d_0\dot{\phi}(t)+\Omega^2_0\phi (t)
+\Omega^2_0\int^t_0 ds\; H(t-s)\dot\phi(s)\; =\; 0 \label{t:r} \\
&&\dot{\psi}(t)+\gamma_v\psi (t)+\int^t_0 ds\;
G(t-s)\psi(s)\; =\; 0, \label{t:n}
\end{eqnarray}
where
\eq
H(t)=c_1\psi (t)\phi (t) +c_2\phi^2 (t),~~~~~
G(t)=-\widetilde{\xi}\;\ddot{\phi}(t)\psi (t) ,
\label{HG}
\en
with
the initial conditions $
\phi (0)=\psi (0)=1$, $\dot{\phi}(0)=0$, and the definition
$\widetilde{\xi}\equiv (\frac{q}{\Lambda})^2 \xi$.
In the following, we integrate them numerically.
We fix $d_0 =\Omega^2_0 =1$ and study the
relaxation of the system at fixed wavenumber $q/\Lambda
=0.1$ as $\gamma_v$ and the MCT coefficients $c_1, c_2$ change.
We use the parametrization Eq.~(\ref{c1212}) for
$c_1, c_2$ and the value of $\xi$ determined from 
Eq.~(\ref{lxi}).

We find that the system depends in a crucial way on
the value of $\gamma_v$. When the value of $\gamma_v$ is
large enough, we expect that the defects quickly diffuse
away and only the density fluctuations are important. It
is clearly seen in the numerical integration of Eqs.~(\ref{t:r}),
(\ref{t:n}) for fixed $c_1$, $c_2$ and $\xi$ in Fig.~5.
The defect auto-correlation
function $\psi (t)$ decays exponentially for large $\gamma_v$.
Thus, in the large $\gamma_v$ limit,
the model essentially 
reduces to the one originally considered by Leutheusser
\cite{leut} where only the $c_2$ term is present.
For smaller values of $\gamma_v$, we expect that the 
system goes into the regime where the result of the previous
section applies such that the slow dynamics of
defects and the coupling between 
defects and density fluctuations play an important role
in stretched dynamics. We can see from Figs.~5 
and 6 that,
as the value of $\gamma_v$ is 
decreased, $\psi (t)$ and $\phi (t)$ are more and
more stretched. We can not, however, take $\gamma_v$ to zero,
since for small enough $\gamma_v$ such that $\gamma_v <
\gamma^c_v$ for some $\gamma^c_v$, $\psi (t)$ starts to 
increase with time and consequently the model becomes unphysical.
This surprising result restricts the value of the parameter 
$\gamma_v$ to $\gamma_v > \gamma^c_v$. It is clear that
$\psi (t)$ and $\phi (t)$ are stretched most when
$\gamma_v$ is very close to but still larger than $\gamma^c_v$.
In fact, the time scale for $\psi (t)$ becomes
extremely large as $\gamma_v\rightarrow\gamma^c_v$
so that it seems to approach some plateau
value $g$ as $t\rightarrow\infty$.
We will discuss later that only when $\gamma_v\simeq\gamma^c_v$ is
the time scale of $\psi (t)$ much greater than that of $\phi (t)$,
$\psi (t)$ is more stretched than $\phi (t)$ and the picture is 
consistent with the basic picture obtained in the previous section.

In adddition to $\gamma_v$, the system also depends on
the MCT coefficients $c_1$, $c_2$ such that as one approaches
the critical surface in the $(c_1, c_2)$ space, the
slowing down of $\phi (t)$
is obtained. In general, we expect that due to the defect 
dynamics, the explicit form of the critical surface will be
different from the one given by Eq.~(\ref{c1212})
with $\sigma_0 =0$. In Fig.~7, the relaxation of $\phi (t)$
is shown when we change $c_1$ and $c_2$ but adjust $\gamma_v$
such that $\gamma_v\simeq\gamma^c_v$ for each case.
We note that in the previous case where $\psi (t)\equiv 1$,
the critical slowing down of $\phi (t)$ is achieved when
$c_1 (\lambda )=(2\lambda -1)/\lambda^2$, 
$c_2 (\lambda )=1/\lambda^2$. As can be seen
from Fig.~7, for the case of $\lambda =0.6$, 
the dynamics of defects
influence the system such that at $(c_1 (0.6), c_2 (0.6))=(0.556,
2.778)$, $\phi (t)$ still relaxes with finite time scales. The
critical slowing down similar to the one with $\psi (t)\equiv 1$
and $(c_1 , c_2)=(0.556,
2.778)$ occurs at $(c_1 ,c_2 )\simeq (0.590,2.714)$ with
larger $c_1$ and smaller $c_2$. We find the similar shift in
the critical surface for $\lambda =0.7$ case as in Fig.~8.
The relaxation at $(c_1 (0.7),c_2 (0.7))=(0.816,2.041)$
for the $\psi (t)\equiv 1$ case is similar to the one at
$(c_1, c_2)=(0.872,1.902)$.

The sequence of the time relaxations of $\phi (t)$ can be most
easily seen in a $\log (-t\dot{\phi}(t))$ versus $\log (t)$
plot. In such plot, the power-law and the von-Schweidler
relaxations are combined to form a minimum with asymptotic
slopes of $-a$ and $b$. The later stage of relaxation with a peak
is well fit by the stretched exponential (Fig.~9). 
As $\gamma_v\rightarrow
\gamma^c_v$, we find that the later stage of the relaxation of
$\psi (t)$ is also well fit by a stretched exponential.
We now use two stretched exponential forms
\eq
\phi (t)=f\;\exp\Bigl( -(t/\tau)^\beta  \Bigr),~~~~~
\psi (t)=g\;\exp\Bigl( -(t/\tau^\prime )^{\beta^\prime} \Bigr)
\label{two}
\en
in Eqs.~(\ref{t:r}) and (\ref{t:n}). 
Since, for $\gamma_v$ near $\gamma^c_v$, the time scale $\tau^\prime$
for $\psi (t)$ is much greater than $\tau$ for $\phi (t)$, we can write
\eq
\exp \Bigl( -(t/\tau^\prime )^{\beta^\prime} \Bigr)\simeq
1-(\frac{\tau}{\tau^\prime})^{\beta^\prime}
(\frac{t}{\tau})^{\beta^\prime}.
\en
We find the following qualitative relations among these parameters from 
Eqs.~(\ref{t:r}) and (\ref{t:n}):
\begin{eqnarray}
&&2^{-\frac{1}{\beta}}=\frac{1-c_1 g}{c_2 f}+\frac{c_1 g}
{c_2 f}\frac{\Gamma (\frac{1+\beta^\prime}{\beta})}
{\Gamma (\frac{1}{\beta})}
(\frac{\tau}{\tau^\prime})^{\beta^\prime}, \label{b:r} \\
&&(\gamma_v -\gamma^c_v )\tau =
\frac{1}{g\;\Gamma (1+\frac{1}{\beta^\prime})}(\frac{\tau}{\tau^\prime})
+fg\beta^\prime \widetilde{\xi}\Gamma (\frac{\beta+\beta^\prime -1}
{\beta})(\frac{\tau}{\tau^\prime})^{\beta^\prime} \label{b:n}
\end{eqnarray}
Numerically, we find that the second term in the right
hand side of Eq.~(\ref{b:n}) is much smaller than the first term.
Thus, we can rewrite Eq.~(\ref{b:n}) as
\eq
(\gamma_v -\gamma^c_v )\tau^\prime\simeq
\frac{1}{g\;\Gamma (1+\frac{1}{\beta^\prime})},
\en
which means that $\tau^\prime$ diverges as $(\gamma_v -\gamma^c_v )^
{-1}$ when $\gamma_v\rightarrow\gamma^c_v$. By obtaining the parameters
$\beta$, $\beta^\prime$, $\tau$ and
$\tau^\prime$ from explicit curve fits of $\phi (t)$
and $\psi (t)$ using Eq.~(\ref{two}) for various values
of $\gamma_v$ near $\gamma^c_v$, we find 
some evidence for this relation (Fig.~10).
Without the second term in the right hand side of Eq.~(\ref{b:r}),
it gives the expression for $\beta$ of the conventional MCT with $c_1$
and $c_2$ ($g=1$): $\beta=-\ln (2)/\ln ((1-c_1)/c_2 f)$.
However, in the presence of defect dynamics, Eq.~(\ref{b:r})
represents a more complicated expression for $\beta$ which 
depends on $\beta^\prime$ as well as $\gamma_v -\gamma^c_v$. We find
again evidence for this relation from our numerical data (Fig.~11).

\vskip8mm
\epsfxsize=8cm
\centerline{\epsfbox{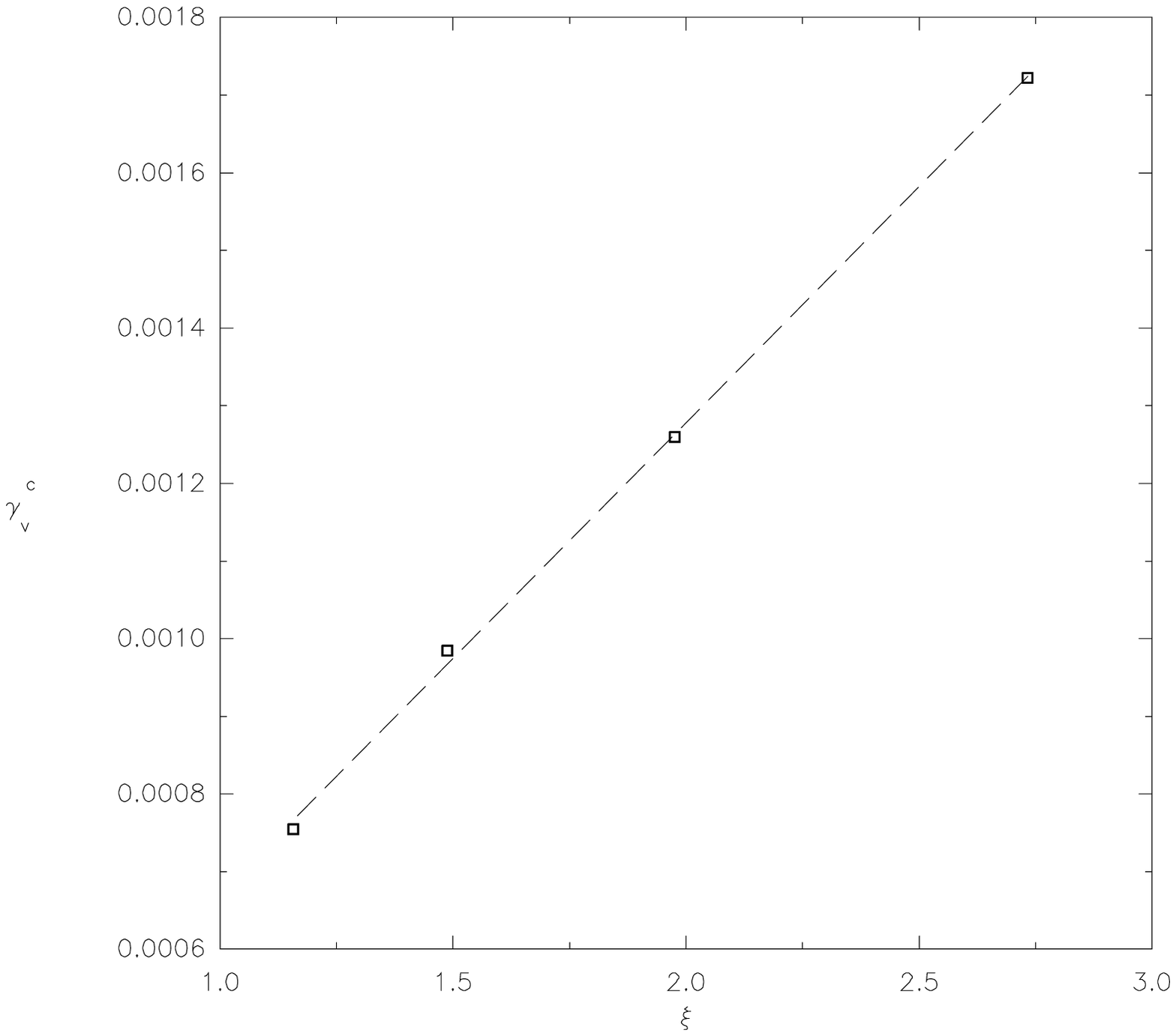}}
\centerline{\small The critical value $\gamma^c_v$ of 
$\gamma_v$ as a function
of the temperature represented by $\xi$ ($\Box$). The dashed
line is a linear fit.}
\centerline{\sc Fig.~12}

It is clear from these results that the separation of time scales
between the defect and the density fluctuations occur only when
$\gamma_v$ is very close to $\gamma^c_v$. 
We find that, as a function of $\xi$, the value of $\gamma^c_v$
shows a smooth linear temperature dependence (Fig.~12).
Under the previous assumption that the slowing down corresponds to the
$\sigma\rightarrow 0$ (or $\gamma_v\rightarrow 0$) limit, we were
able to determine the temperature dependence of the metastability
parameters except for that of $\sigma$ itself. This was done
by assuming that the system arranges itself to be on the 
critical surface. The present analysis of the model including
the defect dynamics determines the temperature dependence
of the parameter $\gamma_v\equiv\sigma y\widetilde{\Gamma}_v q^2$
when the system is on the critical surface.

\section{Discussion}
\label{sec:cut}

It was discovered by Das and Mazenko \cite{dm} that the nonhydrodynamic
correction due to $\Sigma_{\hat{V}\rho}({\bf q},\omega )$
included
in the representation like Eq.~(\ref{original:r})
cuts off the sharp nature of the ergodic-nonergodic 
transition. In the presence of the new variable $n$, however,
we have a very complicated expression, Eq.~(\ref{G_rhr}) 
for the density response function. In order to see
the cutoff effect, let the viscosity $\Gamma$
become arbitrarily large. Eq.~(\ref{G_rhr}) then 
reduces to
\eq
G_{\rho\hat{\rho}}\QO\simeq
\frac{\omega}{
\omega (\omega +iq\Sigma_{\widehat{V}\rho}\QO)
+iq\,\Sigma_{\widehat{V}n}\QO\,\Sigma_{\widehat{n}\rho}\QO
}. \label{cut:Grr}
\en
If the self-energies $\Sigma_{\hat{V}n}$
and $\Sigma_{\widehat{n}\rho}$, as well as $\Sigma_
{\widehat{V}\rho}$ are set to zero, $G_{\rho\widehat{\rho}}(t)$ 
reaches a finite value as time $t\rightarrow\infty$.
The presence of these 
self-energies makes $G_{\rho\widehat{\rho}}(t)$
decay slowly and thus provides the cutoff effect.
The evaluation of these self-energies in principle can be done
to yield complicated expressions in terms of $\phi (t)$
and $\psi (t)$. The analysis including this effect 
thus can be performed but will
be a very difficult task even numerically. 

Throughtout this paper, we have concentrated on the time relaxation 
behavior of the system near the glass transition. At this stage, it
is important to note the implication of our model on the
temperature dependence of the viscosity: $\eta (T)$. Traditionally,
many different expressions have been used to fit the experimentally
observed $\eta (T)$. These include the Arrhenius form $\sim
\exp (A/T)$, the Vogel-Fulcher form $\sim\exp (B/(T-T_{VF} ))$ and
the power-law $\sim |T-T_0 |^{-\gamma}$. These forms are able to fit
the experimental data only over limited temperature ranges.
In MCT, the temperature dependence of the viscosity is
given by \cite{mct:gen}
\eq
\eta (T)\sim\tau_{\alpha} (T)\sim |\sigma_0 |^{-\gamma },
\label{etat}
\en
where $
\gamma =\frac{1}{2a}+\frac{1}{2b}$.
According to conventional MCT, $\sigma_0$ is assumed
to be proportional to $T_0 -T$ and $\gamma$ is temperature independent.
Thus, the conventional MCT predicts that the viscosity shows
a power-law divergence as the temperature approaches $T_0$.
In our model, the situation is quite different. The temperature
dependence of $\sigma_0$ is given by Eq.~(\ref{sigsig0}) without
any special temperature. More importantly, $\gamma=1/(2a)+1/(2b)$
in this model is a function of temperature which is represented 
by the parameter $\xi$. Thus we have
\eq
\eta (\xi )=|f(\xi )\sigma (\xi ) |^{-\gamma (\xi )},
\label{etaxi}
\en
where $f(\xi )=\frac{2}{3}[3y+2C]\xi\lambda(1-\lambda )^2$.
Therefore, the temperature dependence of the viscosity is
governed mainly by the behavior of the exponent 
$\gamma (\xi )$ in Eq.~(\ref{etaxi}) as a function of $\xi$, 
{\it i.e.}
$\log\eta (\xi )\sim\gamma (\xi )$. As shown in Fig.~13, 
$\gamma (\xi )$
increases as the temperature decreases.
We do not expect, however, that Eq.~(\ref{etaxi}) can be used directly
to fit the experimental data, since $a$ and $b$
as functions of $\xi$ are model-dependent. 
Especially the rapid increase of $\log\eta (\xi )$ near the lower bound
of $\xi$ should not be understood as a physical result as
noted in Sec.~\ref{sec:cnn1}.
But this analysis clearly indicates that
our model is consistent with the generic feature
of the observed $\eta (T)$ which increases with
decreasing temperature.
\epsfxsize=8cm
\centerline{\epsfbox{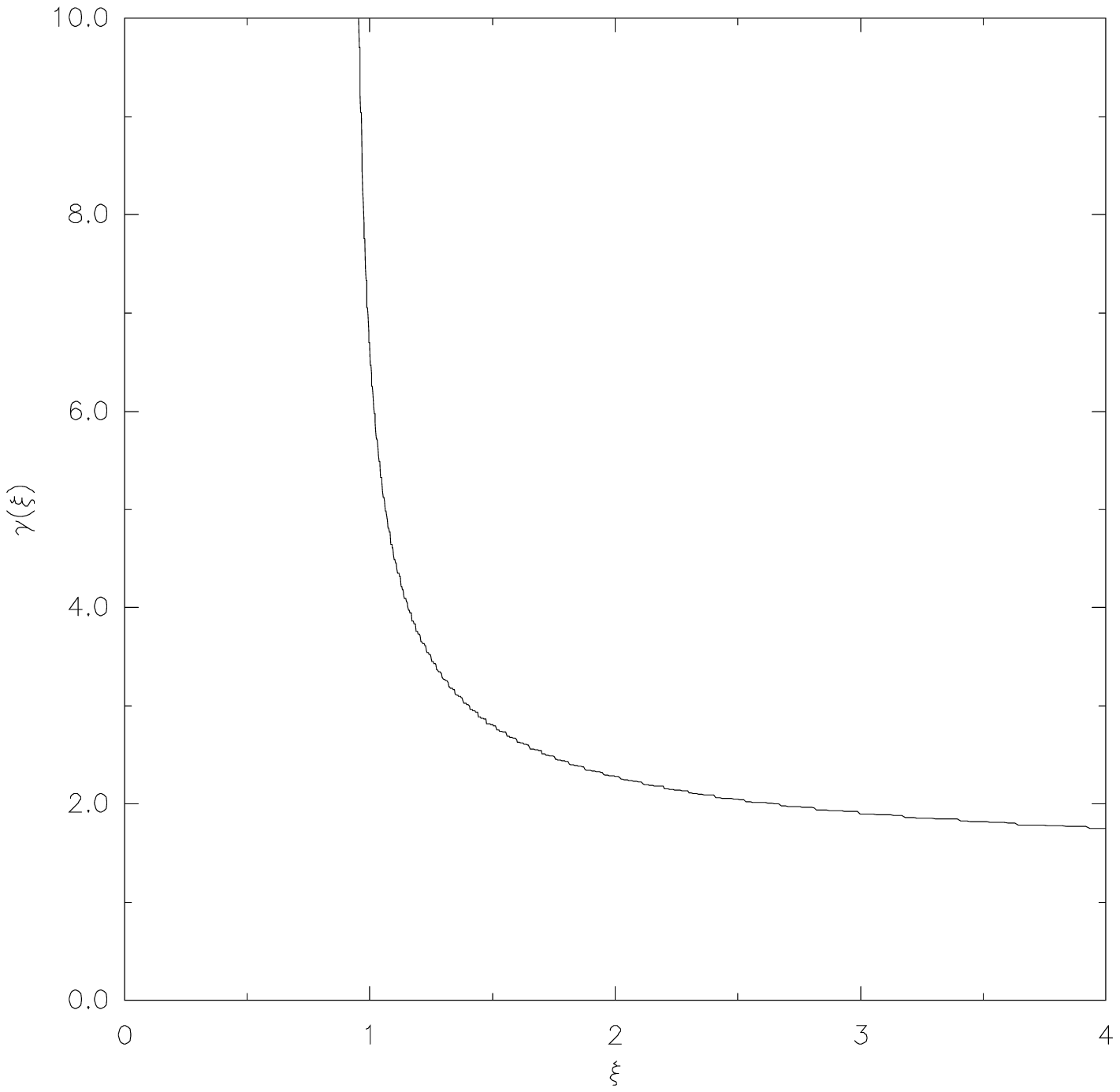}}
\centerline{\small The parameter $\gamma =1/(2a)+1/(2b)$ as a function  
of $\xi$}
\centerline{\sc Fig.~13}
%

%%%%%%%%%%%%%%%%% Appendix %%%%%%%%%%%%%%%%%%%%%%%%%%%%%%%%
\appendix
\section*{ 
The model with a general potential $\lowercase{h(n)}$}
In the present work, we have used an explicit double-well parametrization 
given by Eq.~(\ref{hhnn}) for the potential energy $h(n)$.
In this appendix, we show that the similar analysis can be 
carried out for a general potential $h(n)$ without changing
any physical results to the lowest order of the perturbation theory.
The only information we use for $h(n)$ is that there exists a
metastable defect density $\bar{n}$ such that $h^\prime (\bar{n})=0$.
Then the potential energy $h(n)$ is described by the following
dimensionless parameters:
\eq
\mu_2 \equiv\frac{h^{\prime\prime}(\bar{n})\bar{n}^2}{A\rho^2_0},~~~~~~
\mu_3 \equiv\frac{h^{\prime\prime\prime}(\bar{n})\bar{n}^3}{A\rho^2_0},
\label{munu}
\en
and in general, $\mu_k \equiv h^{(k)}(\bar{n})\bar{n}^k/(A\rho^2_0 )$,
$k=2,3,\cdots$.
Thus, the double-well potential given by Eq.~(\ref{hhnn})
is a particular case where
\eq
\mu_2 =\sigma y ,~~~~\mu_3 =2y(1+\sigma ),~~~~\mu_4 =6y,
\label{hn:mu}
\en
and $\mu_k =0$ for $k=5,6,\cdots$.

To the one-loop order, however, only two
parameters, $\mu_2$ and $\mu_3$ are 
relevant, so that the  
coefficients $d_1$ and $d_2$ in Eq.~(\ref{d12:0}) 
are completely described by $x$, $\mu_2$ and $\mu_3$.
The parameters $\mu_k$, $k=4,5,\cdots$ generate the 
terms in the action, Eq.~(\ref{S}) that contribute to the 
two-loop and higher order diagrams. 
This is basically the reason
why the physical situation at the one-loop order
is well described by the
double-well potential, which has two independent parameters $\sigma$
and $y$ beside $\bar{n}$.  
 
We note that Eq.~(\ref{s00}) was the self-consistent condition
that leads to the separation of the time scales between the density
and the defect variables. For a general potential
it is given by
\eq
\left|\frac{(\mu_2 +x)^2}{\mu_2}\right|\ll 1.
\label{s00:gen}
\en
In terms of $\mu_2$, $\mu_3$ and $x$,
the coefficients $d_1$ and $d_2$
are given by
\begin{eqnarray}
&&d_1 (x,\mu_2 ,\mu_3 )=-\frac{\mu^2_3 x}{\mu^2_2}(2+\frac{x}{\mu_2})  
-2\Bigl[ 2(\mu_2 +\mu_3 )+\frac{\mu^2_3 x^2}{\mu^2_2}\Bigr]
(1+\frac{x}{\mu_2})^2 \label{l1:gen} \\
&&d_2 (x, \mu_2 ,\mu_3 )=1-2\frac{\mu_3 x}{\mu_2}(1+\frac{x}{\mu_2}) 
+\Bigl[ 2(\mu_2 +\mu_3 )+\frac{\mu^2_3 x^2}{\mu^2_2}\Bigr]
(1+\frac{x}{\mu_2})^2 \label{l2:gen}
\end{eqnarray}
As in the case of the double-well potential, a solution to the
critical condition given by Eq.~(\ref{d12:cr}) 
which is consistent with Eq.~(\ref{s00:gen}) exists only when we
take
\eq
x\rightarrow 0,~~~~\mu_2\rightarrow 0,~~~~\frac{x}{\mu^2_2}
\rightarrow \widetilde{C},
\en
for some $\widetilde{C}$. This limit corresponds to
the case where the coupling is weak and the potential
gets very flat near the metastable defect density
$\bar{n}$. We note that $\mu_3$
must be finite, otherwise the model reduces to the
Leutheusser model \cite{leut}, where there is no
stretching. This indicates that the potential
is at least a cubic or higher order polynomial 
in $(n-\bar{n})$ near $\bar{n}$ in order to have the stretched
exponential relaxation. In this limit, we have
\begin{eqnarray}
&&d_1 (x,\mu_2 ,\mu_3 )=-2\mu^2_3\widetilde{C}-4\mu_3-
(4+8\mu_3\widetilde{C}+\mu^2_3\widetilde{C}^2)\mu_2 +{\cal O}(\mu^2_2 )\\
&&d_2 (x,\mu_2 ,\mu_3 )=1+2\mu_3+(2+2\mu_3\widetilde{C})\mu_2
+{\cal O}(\mu^2_2 )
\end{eqnarray}
As done in Sec.~\ref{sec:cnn1},
we can express the metastability
parameters in terms of the temperature $\xi$:
\eq
\mu_3 =\frac{1}{2}(\frac{1}{\xi\lambda^2}-1),~~~~~
\widetilde{C}=\frac{2\Bigl( 2-\frac{2\lambda +1}{\xi\lambda^2}\Bigr)}
{\Bigl( \frac{1}{\xi\lambda^2}-1 \Bigr) ^2},
\en
\eq
\xi=\frac{1}{\lambda^2}\Bigl[ 1-\frac{3(2\lambda -1)}
{2 [2\lambda +1 +\sqrt{4\lambda^2 +10\lambda +13}]   }
\Bigr] .
\en
Because of the particular parametrization of $\mu_2$ and
$\mu_3$ in terms of $\sigma $ and $y$,
the explicit form for the temperature dependence of the parameters
for the double-well potential
is slightly different from that of the general potential.
(See Fig.~14.) But, any of the physical pictures
change from the the double-well potential case. Thus, the double-well 
parametrization is a very good approximation
to the lowest order that provides a concrete physical picture
that the observed slowing down corresponds to the weak coupling
between the density and the defect variables and the low activation 
barrier for the metastable defect.
\vspace{1.5\baselineskip}

\vskip8mm
\epsfxsize=8cm
\centerline{\epsfbox{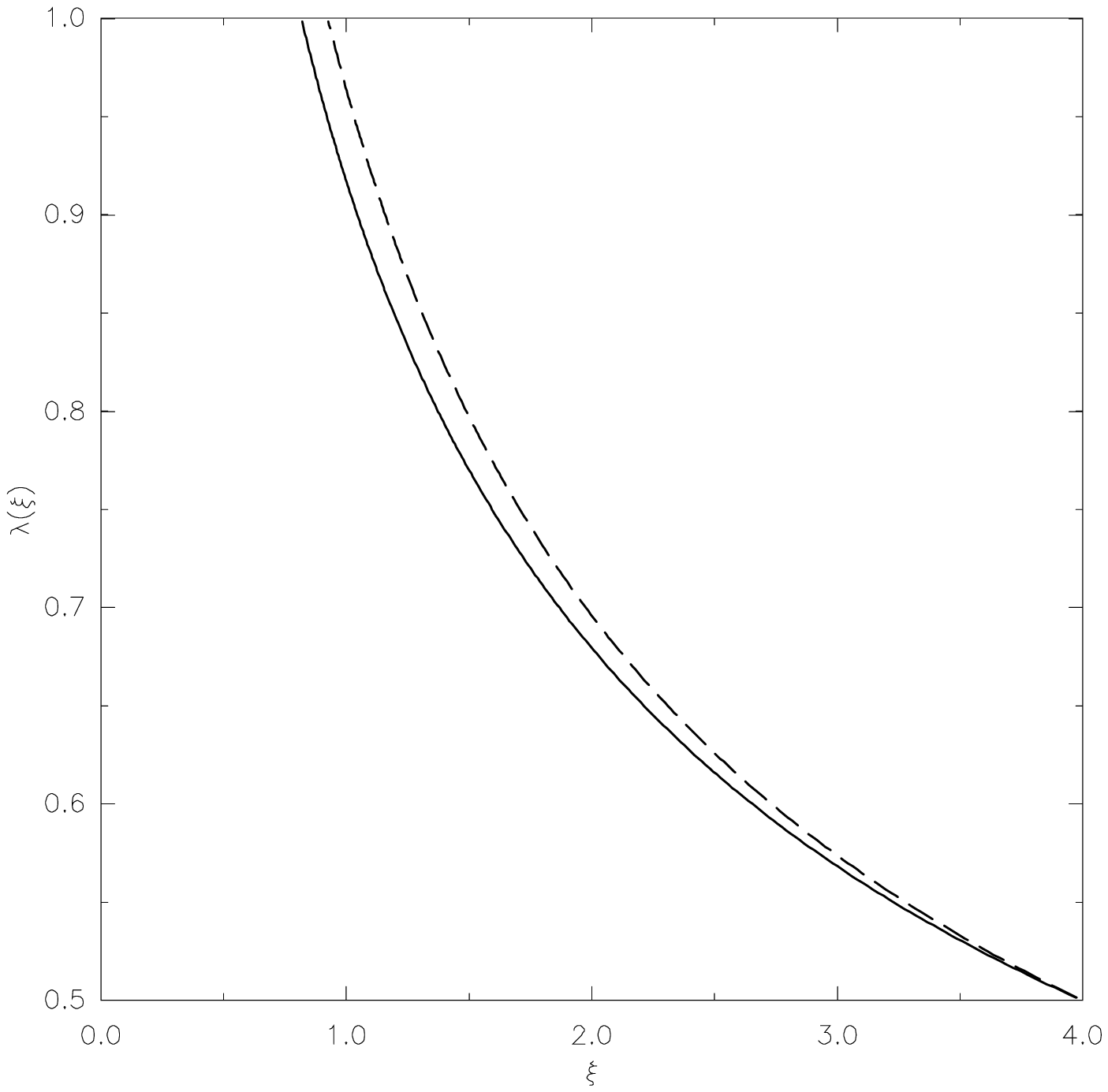}}
\centerline{\small The parameter $\lambda$ as a function  
of $\xi$ for a general (solid) and the double-well (dashed) potentials}
\centerline{\sc Fig.~14}
\vspace{\baselineskip}

\centerline{\large\bf Acknowledgements}
\vspace{0.75\baselineskip}

This work was supported by the National Science Foundation Materials
Research Laboratory at the University of Chicago.

%%%%%%%%%%%%%%%%%%%%%% REFERENCES %%%%%%%%%%%%%%%%%%%%%%%%%%%%%%%

%%%%%%%%%%%%%%%%%%% FIGURES %%%%%%%%%%%%%%%%%%%%%%%%%%%%%%%%%%%%%
\newpage
\centerline{\large\bf Figure Captions}
\vspace{\baselineskip}

\noindent Fig.~1.  {\small
A schematic plot of the sequence of relaxation behaviors predicted
by MCT; (a) power-law decay relaxation $f+A_1 t^{-a}$; (b)
von-Schweidler relaxation $f-A_2 t^b$; (c) primary relaxation
$A_3 e^{-(t/\tau )^\beta }$; and (d) exponential relaxation
$e^{-\gamma t}$. }
\vspace{1.5\baselineskip}

\noindent Fig.~2.  {\small
One-loop diagrams contributing to $\Sigma_
{\widehat{g}\widehat{g}}\QO$. }
\vspace{1.5\baselineskip}

\noindent Fig.~3.  {\small
One-loop diagrams contributing to
$\Sigma_{\widehat{n}\widehat{n}}\QO$.  }
\vspace{1.5\baselineskip}

\noindent Fig.~4.  {\small
The exponents $a$ and $b$ as functions of temperature
represented by $\xi$. }
\vspace{1.5\baselineskip}

\noindent Fig.~5. {\small
The defect auto-correlation function $\psi (t)$
for fixed $(c_1 ,c_2)=(0.556,2.778)$ and for various
$\gamma_v$: (a) $2.0\times 10^{-3}$, (b) $1.8\times 10^{-3}$,
(c) $1.75\times 10^{-3}$, (d) $1.725\times 10^{-3}$,
(e) $1.7225\times 10^{-3}$, (f) $1.7215\times 10^{-3}$,
(g) $1.721\times 10^{-3}$, (h) $1.72\times 10^{-3}$. 
We can estimate the value of $\gamma^c_v$ as $1.72\times 10^{-3}<
\gamma^c_v < 1.721\times 10^{-3}$. }
\vspace{1.5\baselineskip}

\noindent Fig.~6.  {\small
The density auto-correlation function $\phi (t)$ 
for fixed $(c_1 ,c_2)=(0.556,2.778)$ and for various
$\gamma_v$; (a) $2.0\times 10^{-3}$, (b) $1.8\times 10^{-3}$,
(c) $1.75\times 10^{-3}$, (d) $1.725\times 10^{-3}$,
(e) $1.7225\times 10^{-3}$, (f) $1.7215\times 10^{-3}$,
(g) $1.721\times 10^{-3}$, and (h) $1.72\times 10^{-3}$. }
\vspace{1.5\baselineskip}

\noindent Fig.~7.  {\small
The density auto-correlation function $\phi (t)$ for various $(c_1 ,
c_2 )$; (a) (0.467,2.944), (b) (0.488,2.905), (c) (0.522,2.842),
(d) (0.556,2.778), and (e) (0.590,2.714). $\gamma_v$ is adjusted for each
case such that $\gamma_v\simeq\gamma^c_v$. (f) $\phi (t)$ at
$(c_1 , c_2 )=(0.556,2.778)$ for the case where $\psi (t)=1$.}
\vspace{1.5\baselineskip}

\noindent Fig.~8.  {\small
The density auto-correlation function $\phi (t)$ for various $(c_1 ,
c_2 )$; (a) (0.719,2.285), (b) (0.779,2.135), (c) (0.816,2.041),
and (d) (0.872,1.902). $\gamma_v$ is adjusted for each
case such that $\gamma_v\simeq\gamma^c_v$. (e) $\phi (t)$ at
$(c_1 , c_2 )=(0.816,2.041)$ for the case where $\psi (t)=1$. }
\vspace{1.5\baselineskip}

\noindent Fig.~9.  {\small
The solid lines are $\log (-t\dot{\phi}(t))$ versus
$\log (t)$ plots for $(c_1 ,c_2 )=(0.467,2.944)$ and for 
(a) $\gamma_v=1.7225\times 10^{-3}$ and (b) $\gamma_v=1.75\times
10^{-3}$. The dashed lines are the stretched exponential fits,
$\phi (t)=f\exp\{-(t/\tau )^\beta \} $, where (a) $f=0.44$, $\beta =
0.804$, $\tau =5859$ and (b) $f=0.44$, $\beta=0.875$, $\tau =4021$.} 
\vspace{1.5\baselineskip}

\noindent Fig.~10.  {\small
The best-fit
values of $[\tau^\prime \Gamma (1+\frac{1}{\beta^\prime})]^{-1}$ for 
various values of $\gamma_v$ ($\Box$). The dashed line is a linear fit.}
\vspace{1.5\baselineskip}

\noindent Fig.~11.  {\small
$2^{-1/\beta}$ versus $\frac{\Gamma (\frac{1+\beta^\prime}{\beta})}
{\Gamma (\frac{1}{\beta})}
(\frac{\tau}{\tau^\prime})^{\beta^\prime}$ plot ($\Box$). 
The dashed line represents 
a linear fit. }
\vspace{1.5\baselineskip}

\noindent Fig.~12.  {\small
The critical value $\gamma^c_v$ of $\gamma_v$ as a function
of the temperature represented by $\xi$ ($\Box$). The dashed
line is a linear fit.}
\vspace{1.5\baselineskip}

\noindent Fig.~13.  {\small
The parameter $\gamma =1/(2a)+1/(2b)$ as a function of the 
temperature represented by $\xi$. }
\vspace{1.5\baselineskip}

\noindent Fig.~14.  {\small
The parameter $\lambda$ as a function  
of $\xi$ for a general (solid) and the double-well (dashed) potentials.}


\begin{references}
\bibitem{fn:1} More generally one must allow coupling between different
wavenumbers. The role of wavenumber will be discussed below.
\bibitem{fn:2}
We could treat more general forms for the potential, but
it does not change our basic conclusions very much.
See Appendix.
\bibitem{fn:3}
Some quartic terms can also generate
the one-loop self-energies. But this contribution is irrelevant to
our present discussion.
In general, quartic and higher order terms in Eq.~(\ref{S})
generate higher-order self-energies.
\end{references}
\end{document}